\documentclass[amsmath,amssymb,aps,citeautoscript,superscriptaddress,twocolumn,showpacs,10pt]{revtex4-1}

\usepackage{bm}
\usepackage{graphicx}
\usepackage{dcolumn}
\usepackage[colorlinks,linkcolor=red,anchorcolor=green,
citecolor=blue,breaklinks]{hyperref}

\usepackage{mathrsfs}
\usepackage{color}
\usepackage{braket}
\usepackage[normalem]{ulem}
\usepackage{multirow}

\graphicspath{{figures/}}
\newcommand{\hA}{\hat{A}}
\newcommand{\hc}{\hat{c}}

\newcommand{\hP}{\hat{P}}
\newcommand{\hH}{\hat{H}}

\newcommand{\hX}{\hat{X}}
\newcommand{\hY}{\hat{Y}}
\newcommand{\hZ}{\hat{Z}}
\newcommand{\hU}{\hat{U}}
\newcommand{\mhU}{\hat{\mathcal{U}}}

\newcommand{\hT}{\hat{T}}
\newcommand{\ot}{\hat{t}}

\usepackage{qcircuit} 
 
\begin{document}

\preprint{}

\title{
Variational Quantum Eigensolver with Reduced Circuit Complexity
}

\author{Yu Zhang}
\email{zhy@lanl.gov}
\affiliation{Theoretical Division, Los Alamos National Laboratory, Los Alamos, NM, 87545, USA}

\author{Lukasz Cincio}
\affiliation{Theoretical Division, Los Alamos National Laboratory, Los Alamos, NM, 87545, USA}

\author{Christian F. A. Negre}
\affiliation{Theoretical Division, Los Alamos National Laboratory, Los Alamos, NM, 87545, USA}

\author{Piotr Czarnik}
\affiliation{Theoretical Division, Los Alamos National Laboratory, Los Alamos, NM, 87545, USA}

\author{Patrick Coles}
\affiliation{Theoretical Division, Los Alamos National Laboratory, Los Alamos, NM, 87545, USA}

\author{Petr M. Anisimov}
\affiliation{Accelerators and Electrodynamics Group, Los Alamos National Laboratory, Los Alamos, New Mexico 87545, USA}

\author{Susan M. Mniszewski}
\affiliation{Computer, Computational and Statistical Sciences Division, Los Alamos National Laboratory, Los Alamos, New Mexico 87545, USA}

\author{Sergei Tretiak}
\affiliation{Theoretical Division, Los Alamos National Laboratory, Los Alamos, NM, 87545, USA}
\affiliation{Center for Integrated Nanotechnologies, Los Alamos National Laboratory, Los Alamos, New Mexico 87545, USA}

\author{Pavel A. Dub}
\affiliation{Chemistry Division, 
Los Alamos National Laboratory, Los Alamos, NM, 87545, USA}

\date{\today}

\begin{abstract}
The variational quantum eigensolver (VQE) is one of the most promising algorithms to find eigenvalues and eigenvectors of a given Hamiltonian on noisy intermediate-scale quantum (NISQ) devices. A particular application is to obtain ground or excited states of molecules. The practical realization is currently limited by the complexity of quantum circuits. Here we present a novel approach to reduce quantum circuit complexity in VQE for electronic structure calculations. Our algorithm, called ClusterVQE, splits the initial qubit space into subspaces (qubit clusters) which are further distributed on individual (shallower) quantum circuits. The clusters are obtained based on quantum mutual information reflecting maximal entanglement between qubits, whereas entanglement between different clusters is taken into account via a new “dressed” Hamiltonian. ClusterVQE therefore allows exact simulation of the problem by using fewer qubits and shallower circuit depth compared to standard VQE at the cost of additional classical resources. In addition, a new gradient measurement method without using an ancillary qubit is also developed in this work. Proof-of-principle demonstrations are presented for several molecular systems based on quantum simulators as well as an IBM quantum device with accompanying error mitigation. The efficiency of the new algorithm is comparable to or even improved over qubit-ADAPT-VQE and iterative Qubit Coupled Cluster (iQCC), state-of-the-art circuit-efficient VQE methods to obtain variational ground state energies of molecules on NISQ hardware. Above all, the new ClusterVQE algorithm simultaneously reduces the number of qubits and circuit depth, making it a potential leader for quantum chemistry simulations on NISQ devices.
\end{abstract}

\maketitle

\section{Introduction}  
One of the major goals of computational chemistry is the development of methods and algorithms for the calculation of the molecular electronic ground and excited state energies and corresponding wave functions from first-principles. Such eigenvalues and eigenvectors can be obtained from the solution of the time-independent electronic Schr\"odinger equation. However, since the invention of classical digital computers in the early 1940s, the exact numerical solution of this central quantum mechanical equation remains infeasible for systems having more than 12 electrons distributed on 184 spin-orbitals~\cite{1559974}. The reason is that the solution space in this approximation (i.e., the Fock space) grows factorially with the system size (e.g., the number of electrons and basis functions). One of the most promising and immediate applications of quantum computers is solving classically intractable quantum chemistry problems~\cite{cao2018quantum,mcardle2020quantum}. 
   
The quantum phase estimation (QPE) algorithm \cite{kitaev1995quantum} represents the natural translation of the full configuration interaction (FCI) procedure to quantum computers \cite{Aspuru-Guzik1704}. However, QPE requires millions of qubits and quantum gates even for relatively small systems, making the algorithm unsuitable for practical applications on near-term noisy intermediate-scale quantum (NISQ) devices ~\cite{preskil2018}. Full quantum eigensolver (FQE), in which the complexity of basic gate operators is polylogarithmic in the number of spin-orbitals, represents another fully quantum algorithm to simulate molecular systems \cite{RN129}. However, FQE might not be suitable for advantageous quantum chemistry demonstrations due to the limitations of NISQ devices. Indeed, leading digital quantum computers based on superconducting qubit technology have limited coherence times ($\sim$~100~$\mu$s) and gate error rates ($\sim$~2 $\times 10^{-2}$ for a two-qubit gate), limiting the possible number of operations that can be executed to evolve the quantum state. Under these conditions, leveraging classical resources as much as possible through a hybrid quantum-classical approach seems to be the most promising route toward achieving quantum chemical advantage on NISQ devices \cite{Moll_2018,cerezo2020variational,bharti2021noisy}. 

The variational quantum eigensolver (VQE) is one of the leading hybrid quantum-classical algorithms developed specifically to simulate molecular systems in their ground ~\cite{RN155,McClean_2016,adts.201800182,fedorov2021vqe} and ultimately excited states ~\cite{PhysRevA.95.020501,PhysRevA.95.042308,PhysRevResearch.1.033062,Higgott_2019,Kawai_2020,https://doi.org/10.1002/qua.26352}. 
In a typical VQE setup, a variational parameterized ansatz is used to represent the trial wave function prepared with a quantum circuit and the expectation value of the qubit Hamiltonian is measured. Then, the parameters of the ansatz are iteratively optimized on a classical computer using the Rayleigh-Ritz variational principle. Although VQE simulations of small molecules have been performed on various quantum architectures such as photons~\cite{RN155}, superconducting qubits~\cite{RN748,Kandala2017Nature} and trapped ions~\cite{RN671}, major efforts are needed to scale up this approach to larger molecular systems of chemical interest. Here, albeit not as large as QPE circuits, VQE would also suffer from the large size of the quantum circuits, which coupled to a classical optimization requiring many 
variational parameters, can render calculations intractable. Consequently, the construction of an efficacious ansatz and improved classical optimizer is an active area of research~\cite{bharti2021noisy,cerezo2020variational}.

There are two main approaches for ansatz design for electronic structure calculations. Hardware-efficient ansatzes~\cite{RN748,Kandala2017Nature} are constructed from repeated, dense blocks of a limited set of parameterized gates that are easy to implement on a quantum device. While such an ansatz requires fewer resources on the quantum processor, because of its totally chemically agnostic nature it might face difficulties in the parameter optimization due to the barren-plateau problem ~\cite{RN861,wang2021noiseinduced,RN670,abbas2020power}. Consequently, simulation of large molecular systems with hardware-efficient ansatz
might become practically challenging and even impossible. Chemically-inspired ansatzes,
such as the Unitary Coupled-Cluster Singles and Doubles (UCCSD) ~\cite{BARTLETT1989133,RevModPhys.79.291,jcp4768229,PhysRevA.98.022322,Alan2018QST}, are designed by using the domain knowledge from classical quantum chemistry. The standard version of UCCSD, however, has unfavorable scaling in the number of gates required for larger molecular systems ~\cite{acs.jctc.9b00236}. Consequently, various approaches are being developed to improve on the UCCSD method to obtain shorter circuits or obtain higher than UCCSD accuracy at comparable circuit depth~\cite{fedorov2021vqe,jctc8b00932,jctc.8b01004,jctc9b00963,Xia_2020,permVQE2021}. Among them, ``iterative VQE" algorithms ~\cite{RN170,PRXQuantum.2.020310,jctc9b01084,RN669,Ryabinkin_2021,yordanov2020iterative}, which instead of using a fixed ansatz, construct problem-tailored ansatzes on-the-fly, have gained attention because of their controllable circuit-size.
 
Two notable iterative VQE algorithms to reduce quantum circuit depth are (fermionic and qubit) ADAPT-VQE~\cite{RN170,PRXQuantum.2.020310} and iterative Qubit Coupled Cluster (iQCC)~\cite{jctc9b01084,RN669,Ryabinkin_2021}. Fermionic-ADAPT-VQE employs a predetermined pool of spin-adapted fermionic (generalized) singles and doubles excitation operators from which the ansatz is dynamically constructed ~\cite{RN170}. The more measurement- and circuit-efficient (e.g. less CNOT gates in particular) qubit-ADAPT-VQE uses selected Pauli words (obtained from mapping fermionic operators into Pauli operators) to form the ``qubit" pool, but the algorithm comes with a larger number of ADAPT iterations because of the larger number of variational parameters (more gradient calculations required) ~\cite{PRXQuantum.2.020310}.
IQCC uses arbitrarily shallow quantum circuit depth at the expense of an iterative canonical Hamiltonian dressing technique employing the qubit coupled cluster (QCC) ansatz. The exponential growth of the ``dressed" Hamiltonian with the size of the QCC transformation in an anti-commuting set could be mitigated using the involuntary linear combinations of Paulis technique~\cite{RN669}. But the algorithm is exponential with respect to the dressing steps.
 
Although circuit depth is one of the most critical metrics for NISQ devices. 
The number of physical qubits on the actual quantum chip (circuit width) is the necessary condition that limits the problem that can be solved. Several techniques based on symmetries presented in the Hamiltonian have been used to reduce the number of qubits~\cite{bravyi2017tapering,RN676,doi:10.1063/1.5110682,PhysRevResearch.3.013039}. But these techniques can only reduce a few qubits and have certain symmetry requirements.
In this work, we present the first-ever algorithm which reduces both quantum circuit depth and width in VQE at the cost of additional classical resources. 
Our novel algorithm, called ClusterVQE, uses the mutual information (measure of correlation between 
spin-orbitals~\cite{Rissler2006,HUANG20051}) to group the qubits into different clusters, which can be distributed to different quantum circuits.
The correlation between different clusters is minimized by the mutual information optimization method and is taken into account by building a ``dressed" Hamiltonian. Even shallower circuit depths can be achieved compared to the state-of-the-art qubit-ADAPT-VQE. Consequently, the new algorithm is more robust to noise making it particularly attractive for NISQ devices. Because the correlations between different clusters are minimized, the dressed Hamiltonian protocol only introduces minor additional classical resources requirements compared to the iQCC method. Most importantly, the new ClusterVQE algorithm removes the entanglement between different clusters via the dressed Hamiltonian. Consequently, ClusterVQE can solve the original problem in a much smaller qubit space (clusters), reducing the number of qubits significantly for simulating large molecules on NISQ devices. 

\begin{figure}
  \includegraphics[width=0.5\textwidth]{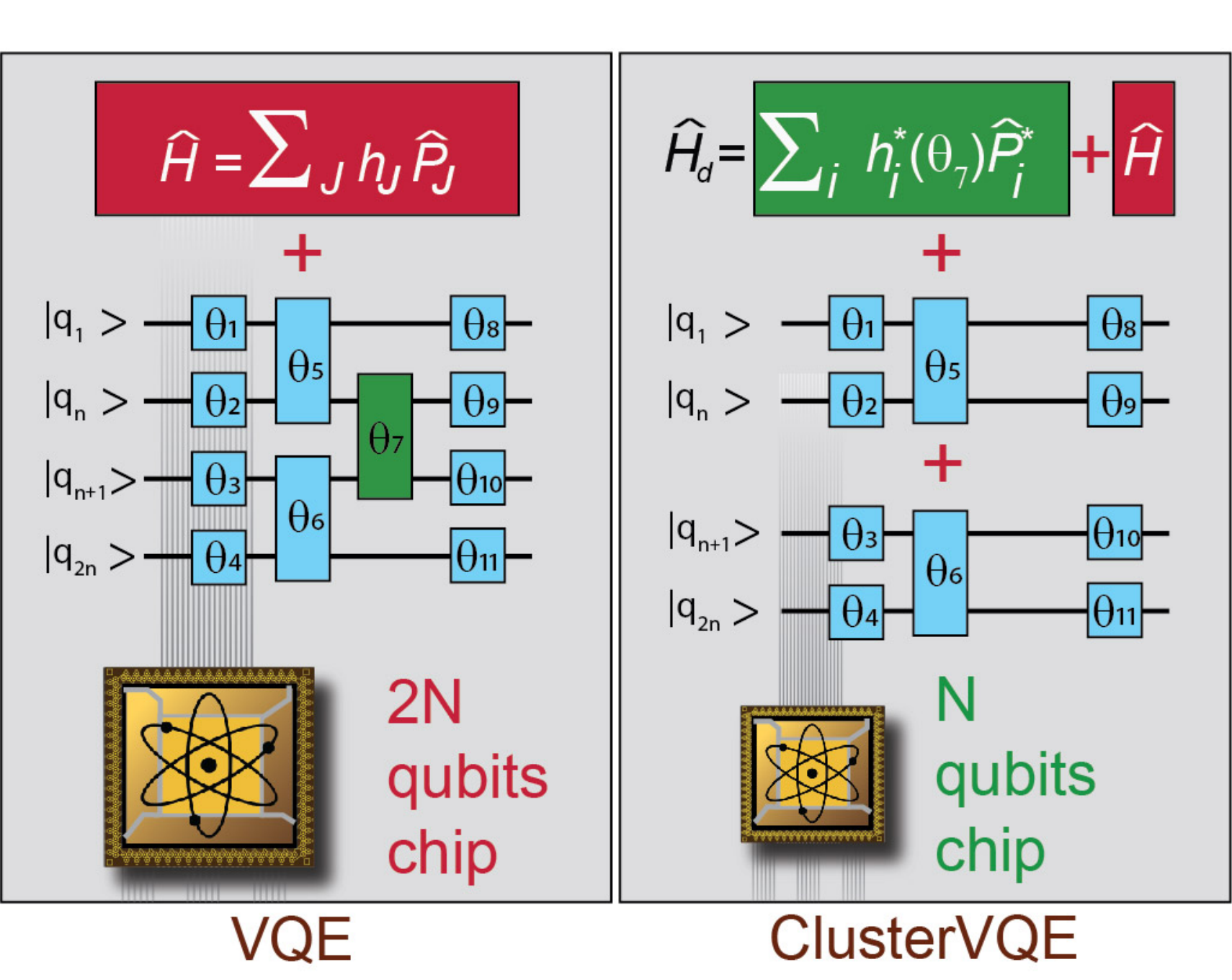}
  \caption{\label{fig1} 
  Finding the lowest eigenvalue of a given qubit Hamiltonian $\hH$ with VQE and ClusterVQE: VQE uses a quantum circuit defined on $m$N qubits ($m=2$ for the example used), whereas ClusterVQE uses N qubits and shorter circuits at the expense of a dressed Hamiltonian $\hH_d$.} 
\end{figure}

\section{Methodology}
\subsection{Mutual Information based Clustering of Qubits}\label{qubtclustering}
The first essential step of ClusterVQE is to split the qubits into different clusters and minimize the correlations between different clusters. With this aim, an entanglement map based on the mutual information (MI) that reflects the electronic correlations among all pairs of qubits is calculated~\cite{permVQE2021}. Previous results obtained for the orbital ordering problem in the Density Matrix Renormalization Group (DMRG) method in classical quantum chemistry calculations~\cite{Rissler2006} showed that the MI is a reliable parameter to quantify the correlation between two quantum particles. Our previous work also demonstrated that an approximate MI is sufficient for the qubit re-arrangement~\cite{permVQE2021} for the reduction of circuit depth. The MI has also been used to construct a compact entangler pool for ADAPT-VQE~\cite{RN854}. According to quantum information theory, the MI between qubits $i$ and $j$ is defined as follows~\cite{Rissler2006}:
\begin{equation}\label{eq_mutual}
    I_{ij}=\frac{1}{2}(S_i+S_j-S_{ij})(1-\delta_{ij}),
\end{equation}
where $S_i$ and $S_{ij}$ are the single- and two-qubit Von Neumann entropies, respectively, and $\delta_{ij}$ is the Kronecker delta. $S_i$ and $S_{ij}$ are obtained from the corresponding one- and two-body density matrix~\cite{permVQE2021}.

Once the MI is obtained, it can be used to cluster the qubits into different groups by minimizing the MI between different clusters which is defined as the summation over MIs between inter-cluster qubits. The qubit clustering can be achieved by using a classical graph partitioning method, such as the graph partition function of the Metis library~\cite{metis} (Appendix.~\ref{app_metis}). Alternatively, this can be mapped into a quadratic unconstrained binary optimization (QUBO) problem and solved on a quantum annealer by either the MI-based clustering method (details are given Appendix~\ref{app_gauge}) or the quantum community detection method (QCD)~\cite{sue_qcd2021} (details are given in Appendix~\ref{qcomm_detection}). Problems with up to 64 variables or nodes can be solved directly on the D-Wave 2000Q quantum annealer. Larger sizes require a quantum-classical approach using \emph{qbsolv}~\cite{qbsolve} + D-Wave.

\subsection{The ClusterVQE algorithm}
Within the VQE algorithm, a parameterized ansatz $\ket{\Phi}=\hU(\theta)\ket{\Phi_r}$ is suggested and encoded on the quantum circuit, where $\ket{\Phi_r}$ is the reference state. In this work, the mean-field Hartree-Fock (HF) state is used as a reference. The unitary operator $\hU(\theta)$ introduces the correlations between qubits. Since current quantum gates can only operate on a few qubits at a time, $\hU(\theta)$ is decomposed as a tensor product of many unitary operators, denoted as entanglers, with each entangler acting on a few qubits, i.e., $\hU(\theta)=\prod^{\mathcal{D}}_j \hU(j)$ where $\hU_j=e^{i\theta_j\hA_j}$ are the entanglers and $\mathcal{D}$ is the number of entanglers in the ansatz. In the widely used UCCSD ansatz, the $\{\hA_j\}$ can be readily obtained by mapping the single/double Fermionic excitation operators into qubit-space as shown in Appendix~\ref{appuccsd}. With the ansatz, the ground-state (GS) energy can be obtained by imposing the variational principle,
\begin{equation}\label{eq_engmin}
E_{GS}=\min_{\theta}\bra{\Phi_r} \hU^\dag(\theta)\hH \hU(\theta)\ket{\Phi_r},
\end{equation}
where the expectation value is measured on a quantum circuit. 
However, the contribution of each entangler to the GS energy is different. Thus, the UCCSD ansatz can be made even more compact by the adaptive construction~\cite{RN170,RN671} in the qubit-ADAPT-VQE, at the cost of more VQE calculations. This algorithm ends up with a compact sequence of entanglers $\hU(\theta)=\prod^{D}_{j=1} \hU_j(\theta_j)$ with $D<\mathcal{D}$, where $\hU_j(\theta_j)=e^{i\theta_j\hP_j}$ and the corresponding $\{\hP_j\}$ is  chosen from the operator pool $\{\hA_j\}$ based on their contributions to the energy weighted by the gradients~\cite{RN170}. 

Alternatively, according to the qubit clustering introduced in Sec.~\ref{qubtclustering}, the entanglers used in the ansatz can be grouped into different sets,
\begin{equation}
\hU(\theta)=
\{\hU_{c_1},\cdots,\hU_{c_M},\hU_{c_{12}},\cdots,\hU_{c_{M-1,M}}\}
\end{equation}
where $M$ is the number of clusters. In this way, the ordering of entanglers in $\hU(\theta)$ is changed and the corresponding parameters can be different as well. The entangler set $\{\hU_{c_i}\}$ only acts on the corresponding cluster, and \{$\hU_{c_{ij}}\}$ represents the entanglers that couple different clusters. 
Consequently, the energy minimization in Eq.~\ref{eq_engmin} can be rewritten as
\begin{eqnarray}\label{eq_edress}
E_{GS}=&&\min_{\theta}\bra{\Phi_r} \prod_i\hU^\dag_{c_i}\left[\prod_{i\neq j}\hU^\dag_{c_{ij}}\hH \hU_{c_{ij}}\right]\hU_{c_i}\ket{\Phi_r} \nonumber\\
=&&\min_{\theta}\bra{\Phi_r} \prod_i\hU^\dag_{c_i}\hH^d \hU_{c_i}\ket{\Phi_r}
\end{eqnarray}
where $\hH_d\equiv\prod_{i\neq j}\hU^\dag_{c_{ij}}\hH \hU_{c_{ij}}$ is the dressed Hamiltonian. Because the $\hH_d$ can be represented in terms of Pauli words and each Pauli word can be trivially rewritten as the tensor product of different clusters, i.e.,  $\hH_d=\sum_k \alpha_k \hP_k=\sum_k \alpha_k \prod_i\hP_k(c_i)$ and the initial state can be decomposed as $\prod_i\Phi_r(c_i)$, the energy in Eq.~\ref{eq_edress} can be further rewritten as
\begin{equation}\label{eq_egs}
    E_{GS}=\min_{\theta}\sum_k\alpha_k\left[ \prod_i\bra{\Phi_r(c_i)}\hU^\dag_{c_i} \hP_k(c_i) \hU_{c_i}\ket{\Phi_r(c_i)}\right].
\end{equation}
Here each expectation value $\bra{\Phi_r(c_i)}\hU^\dag_{c_i} \hP_k(c_i) \hU_{c_i}\ket{\Phi_r(c_i)}$ only involves the qubits in the cluster $i$. Hence, the energy of a larger problem can be measured on $M$ smaller quantum circuits after Hamiltonian dressing, as shown by the schematic diagram in Figure~\ref{fig1}. 
In practice, like the qubit-ADAPT-VQE algorithm, clusterVQE adaptively grows the ansatz in Eq.~\ref{eq_edress}. The ansatz used in the $j^{th}$ iteration is denoted as $\mhU_j$. However, in contrast to the growing ansatz in qubit-ADAPT-VQE~\cite{RN170} and growing Hamiltonian in iQCC~\cite{jctc9b01084}, the ansatz in ClusterVQE is decomposed onto smaller quantum circuits, and only the entanglers between different clusters are used to dress the Hamiltonian. Consequently, the iterative construction breaks the original VQE algorithm with deeper and wider quantum circuits into a series of quantum circuits with a shallower depth and a smaller number of qubits, as illustrated by Figure~\ref{fig1}.

The entangler $\hU_k$ chosen in the ansatz $\mhU_j$ for the $j^{th}$ iteration is determined by its contribution to the energy weighted by the gradient of energy with respect to the parameter, i.e., the $\hU_k$ terms with largest gradients $\frac{\partial E}{\partial \theta_k}$ are selected~\cite{RN170}, where the gradient is defined as 
\begin{eqnarray}\label{eq:derivative}
\frac{\partial E}{\partial\theta_k}=&&\frac{\partial}{\partial \theta_k}\bra{\Phi_r}\hU^\dag_k(\theta_k)\hH_{d,j-1}\hU_k(\theta_k)\ket{\Phi_r} \nonumber|_{\theta_k=0}\\
=&& i\bra{\Phi_r}\hU^\dag_k(\theta_k)[\hH_{d,j-1},\hP_k]\hU_k(\theta_k)\ket{\Phi_r}.
\end{eqnarray}
Where $\hH_{d,j-1}$ is the dressed Hamiltonian of the $j-1$ iteration. It should be noted that $\hH_{d,j}$ may be the same as the $\hH_{d,j-1}$ if the selected entangler of the $j^{th}$ iteration is not used to dress the Hamiltonian. Since the measurement of each gradient term is independent, all the gradients can be measured in parallel with several uncoupled quantum computers~\cite{RN170}. 

\begin{figure}
  \includegraphics[width=0.5\textwidth]{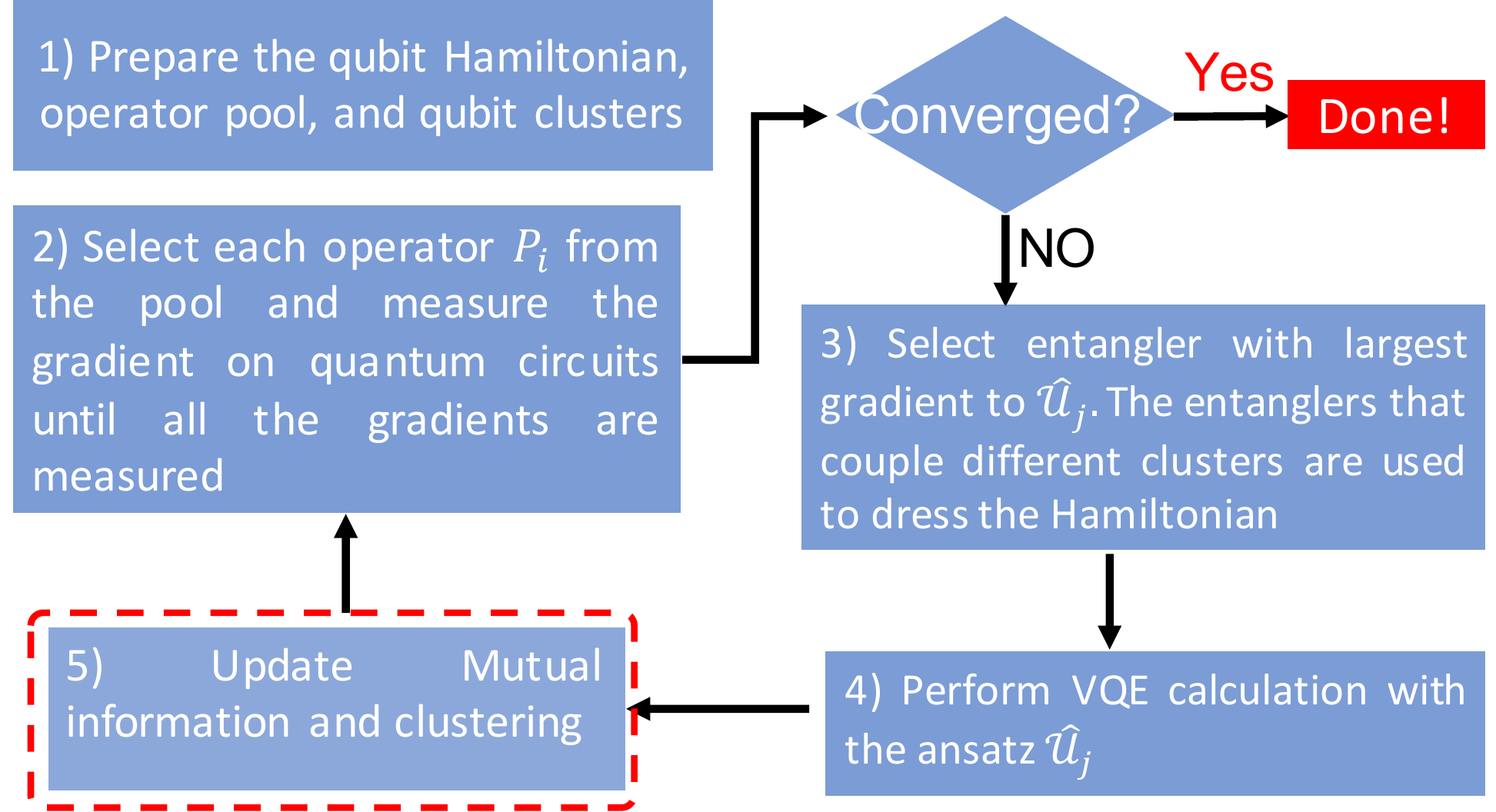}
  \caption{\label{fig_flow} Flowchart of the ClusterVQE algorithm. The fifth step in the red-dashed box is skipped if MI is provided.}
\end{figure}

The flowchart of the ClusterVQE scheme is presented in Figure~\ref{fig_flow}. In the first step,  the qubit Hamiltonian and operator pool are prepared: a) Provided classical mean-field calculation of the molecule, map the second quantized Hamiltonian into its qubit counterpart by using either Jordan-Wigner (JW)~\cite{RN126}, Bravyi-Kitaev (BK)~\cite{BRAVYI2002210}, or parity~\cite{parity2012,bravyi2017tapering}  encoding. b) Generate the UCCSD excitation operators and map them into Pauli words. Here only Pauli words with the different flip indices (Appendix.~\ref{appuccsd}) are chosen and added into the excitation pool $\{\hA_i\}$. In addition, clustering of qubits is performed if MI is provided. Otherwise, the qubits in the first iteration are simply grouped according to their spin index. If the largest gradient in the second step is smaller than the convergence criteria $\epsilon$, the energy minimization criteria are satisfied and the algorithm exits from the iterative loop. The fifth step is only triggered if clustering on-the-fly is desired. 


Since some of the entanglers in $\mhU_j$ are used to dress the Hamiltonian, the circuit depth is smaller than that of qubit-ADAPT-VQE. Besides, because the clustering minimizes the entanglement between different clusters, the number of dressing entanglers is much smaller than that of iQCC. Consequently, ClusterVQE can balance the computational costs between the Quantum Processing Unit (QPU) and CPU. The circuit depth/width is further reduced compared to qubit-ADAPT-VQE, and exponential scaling in growing of the Hamiltonian in iQCC is avoided. As a result, ClusterVQE is more robust to noise, as shown and discussed later, making the ClusterVQE algorithm more suitable for NISQ devices than qubit-ADAPT-VQE. Most importantly, compared to previous approaches, ClusterVQE can solve larger problems on smaller quantum circuits. 

\subsection{Analytical gradients for VQE optimization}
In order to improve the algorithm performance in the presence of noise, an analytical gradient is essential for classical optimization. It should be noted that the gradient for VQE optimization is different from the gradients in Eq.~\ref{eq:derivative} which is used to chose entanglers for each iteration. Ref.~\cite{adts201800182} suggests calculation of the analytical gradients by performing a measurement on an ancillary qubit~\cite{Alan2018QST,Alan201801arxiv}. A similar approach has been developed in this work for the analytical gradient measurements but without introducing an ancillary qubit. The gradient of energy with respect to the variational parameter reads as follows.
\begin{equation}
\frac{\partial E}{\partial\theta_i}=2\text{Im}\left\{\bra{\Phi_r}\hU^\dag \hH\hP^{\text{eff}}_i \hU\ket{\Phi_r}\right\}    
\end{equation}
where $\hP^{\text{eff}}_i$ is the dressed Pauli word as shown in Appendix~\ref{appgradient}. Therefore, the energy gradient can be measured in the same way as the energy measurement by replacing $\hH$ with $\hH\hP^{\text{eff}}_j$ and no ancillary qubit is required. On a noiseless simulator, VQE calculations with numerical gradients or analytical gradients give the same results. However, analytical gradients are much more stable on noisy simulators and quantum computers. 

It should be noted that even though the ClusterVQE method can reduce the number of qubits and ansatz complexity by distributing the measurements on smaller quantum circuits, the number of measurements increases with the number of entanglers used in the dressed Hamiltonian. To further reduce the number of measurements, future work may consider recently developed measurement reduction techniques~\cite{RN818,RN826,RN824,RN817,RN757,RN784,RN626,William2019arxiv}.

\section{Results and discussion}
To demonstrate the validity of the ClusterVQE algorithm, we have simulated several molecules on both quantum simulator and IBM quantum devices. In order to perform the simulations, an in-house modified Qiskit code ~\cite{Qiskit} has been developed. The Jordan-Wigner operator transformation~\cite{RN126} and the Limited-memory BFGS Bound (L-BFGS-B) optimizer~\cite{bfgs} were used in this work. The obtained results are compared with that of exact diagonalization. In order to make the comparison over different methods (VQE, qubit-ADAPT-VQE, iQCC) meaningful, the same ansatz, qubit-UCCSD (details in Appendix~\ref{appuccsd}), is used for all simulations below. The convergence criteria of $\epsilon=10^{-4}$ and $5\times10^{-3}$ are used for LiH and N$_2$, respectively, if not specified elsewhere. For the N$_2$ molecules, the maximum number of iterations is 25, unless convergence occurs earlier. 

\begin{figure}
  \includegraphics[width=0.5\textwidth]{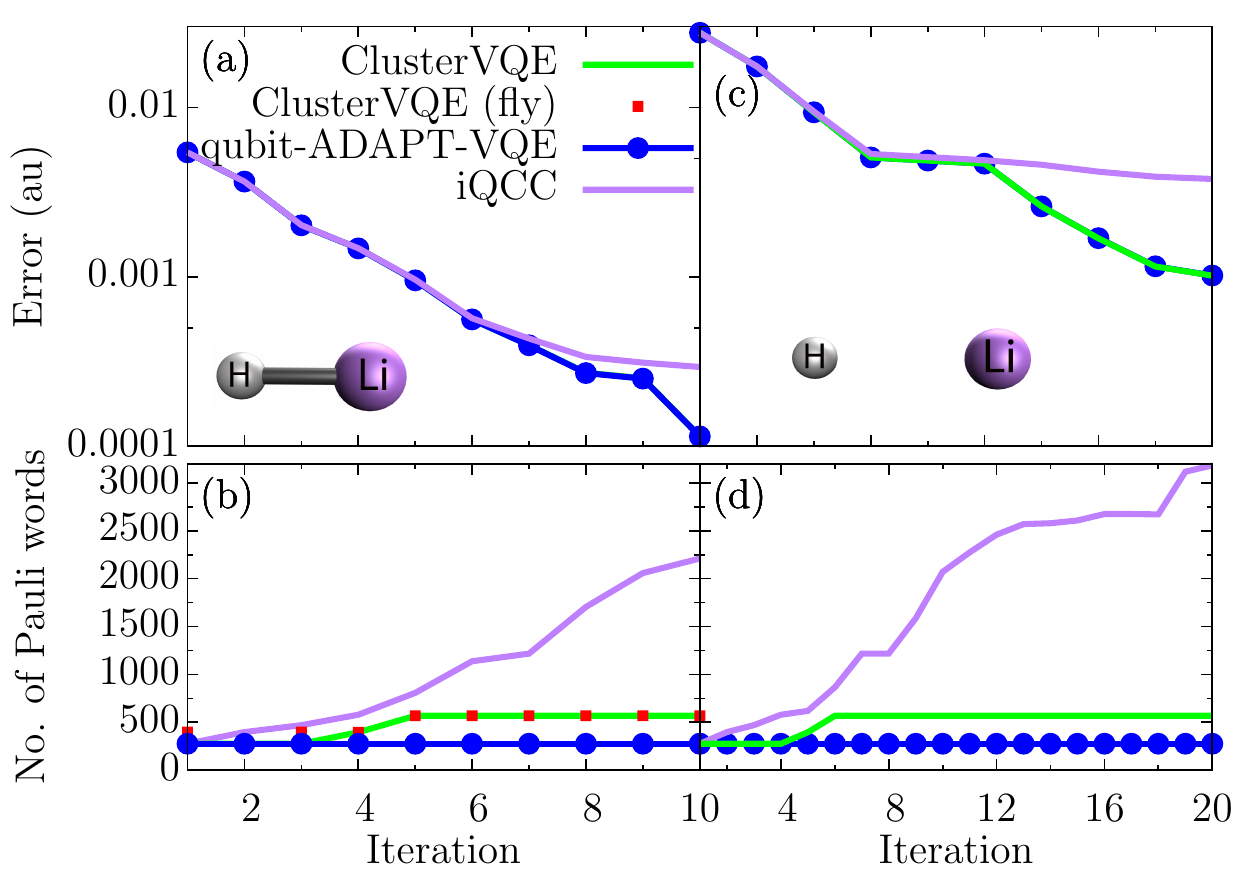}
  \caption{\label{fig_lih} GS Energy errors (a,c) and (b,d) number of Pauli words per each iteration for the LiH molecule with different Li-H bond lengths, 1.547~\AA (a,b) and 2.4~\AA (c,d), obtained from the qubit-ADAPT-VQE, iQCC, and ClusterVQE algorithms.}
\end{figure}

\begin{figure}[h!]
  \includegraphics[width=0.5\textwidth]{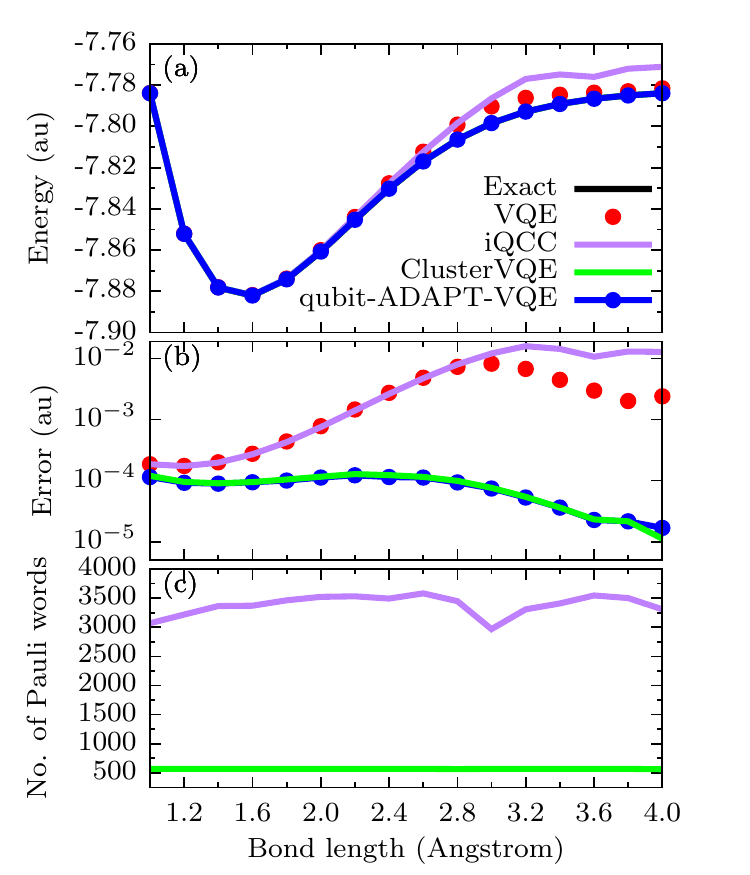}
  \caption{\label{fig_pes} (a) GS energy, (b) energy error, and (c) number of Pauli words in the qubit Hamiltonian of LiH for different bond lengths obtained from different methods. For VQE, one layer of the qUCCSD ansatz is used. One layer of the qUCCSD ansatz is also used as the operator pool for the other methods.}
\end{figure}

\subsection{LiH molecule}
We start with simulating the ground state of the LiH molecule. LiH requires 10 qubits with the minimal STO-3G basis set and frozen core orbitals.
Figures~\ref{fig_lih}(a) and (c) compare the energy convergence of the qubit-ADAPT-VQE, ClusterVQE, and iQCC approaches for different Li-H bond lengths (1.547~\AA and 2.4~\AA). Because iQCC only implements one entangler on the quantum circuit, the number of parameters to be optimized is also smaller compared to that of qubit-ADAPT-VQE and ClusterVQE. Consequently, a smaller number of optimization cycles is required per iteration. For the iQCC method, once the entanglers in the previous iteration are used to dress the Hamiltonian, these parameters are fixed. In contrast, the parameterized ansatz in the qubit-ADAPT-VQE and ClusterVQE approaches are to be re-optimized at each iteration, making them more flexible to achieve better convergence. As a result, the convergence of iQCC is slightly worse than that of qubit-ADAPT-VQE for a given number of iterations,  as shown in Figures~\ref{fig_lih}(a) and (c). For ClusterVQE, because only a few operators are used to dress the Hamiltonian, the performance of ClusterVQE is almost the same as that of qubit-ADAPT-VQE. Because the Hamiltonian of the iQCC and ClusterVQE methods is dressed by the entanglers, the number of Pauli words in the dressed Hamiltonian is increased. However, compared to iQCC, only two operators are used to dress the Hamiltonian in ClusterVQE as shown, the number of Pauli words of the dressed Hamiltonian does not increase significantly. In contrast, the number of Pauli words in iQCC increases exponentially with the number of iterations before saturation, see Figures~\ref{fig_lih}(b) and (d). We also compared performance of the qubit-ADAPT-VQE, iQCC, and ClusterVQE approaches on the ground-state potential energy surface (PES) of LiH as shown in Figure~\ref{fig_pes}. Overall, ClusterVQE can achieve the same energy error as that of qubit-ADAPT-VQE but with smaller quantum circuits. And ClusterVQE uses less Pauli words compared to iQCC, as shown in Figure~\ref{fig_pes}. 

We also compared the performance of the different clustering methods used (Appendix.~\ref{appendix1}). For most geometries of the LiH molecule, these three methods produce the same clustering. However, at the bond length of 2.4~\AA, the Metis graph partition method results in less optimal clustering. Consequently, more entanglers are needed to dress the Hamiltonian and ClusterVQE with the Metis method ends up with 808 Pauli words. This comparison suggests that the QCD and MI-based clustering methods provide better and more stable performance in clustering and help improve the ClusterVQE performance.

\begin{figure}
  \includegraphics[width=0.5\textwidth]{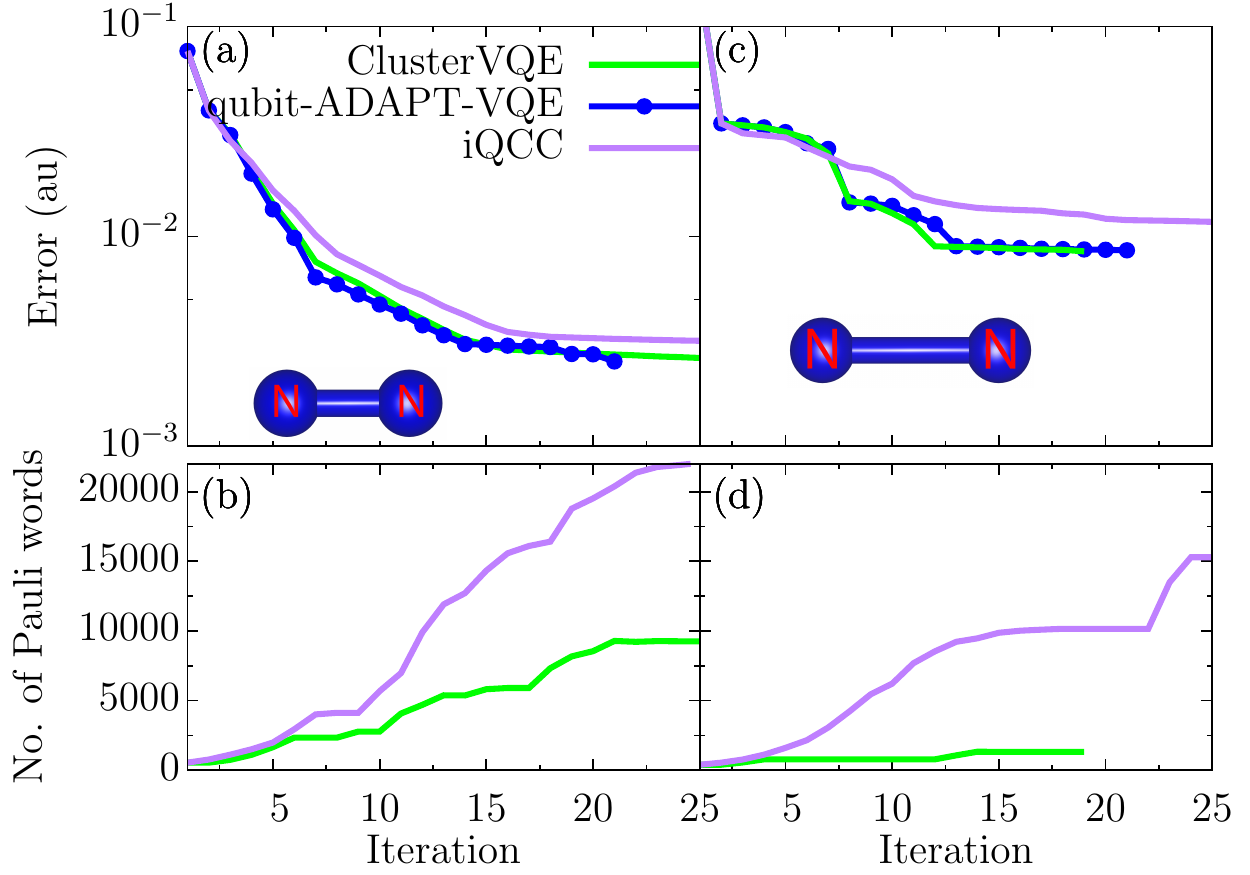}
  \caption{\label{fign2} GS Energy errors (a,c) and number of Pauli words (b,d) per each iteration for the N$_2$ molecule with different bond lengths, 1.09~\AA and 1.6~\AA, obtained from the qubit-ADAPT-VQE (blue), iQCC (purple) and ClusterVQE (green) methods.}
\end{figure}

\subsection{N$_2$ molecule}
We next demonstrated the validity of ClusterVQE in the strongly correlated regime. The N$_2$ molecule is a prototype system for testing multi-reference methods in quantum chemistry due to its strong electronic correlations particularly when the bond is stretched. Here N$_2$ at equilibrium (1.09~\AA) and stretched configurations (1.6~\AA) are used as examples. The 14-qubit N$_2$/STO-3G system is simulated after freezing the lowest 
two spin-orbitals. 
As shown in Figures~\ref{fign2}(a) and (c), the energy convergence of ClusterVQE is almost the same as that of qubit-ADAPT-VQE and is better than that of iQCC. Moreover, the size of the dressed Hamiltonian in ClusterVQE is 
much smaller than that of iQCC. It should be noted that the stretched N$_2$ molecule has stronger correlations, however, accompanied with a better separation between clusters. Consequently, less dressing is required for the stretched N$_2$, and ClusterVQE only slightly increases the size of the dressed Hamiltonian. In contrast, the size of the Hamiltonian exceeds 15,000 Pauli words in iQCC after 25 iterations. These results suggest that, even for strongly correlated systems, our ClusterVQE 
reduces circuit complexity without a significant increase in Hamiltonian size. 

\subsection{Noisy simulation and clustering on-the-fly}
\begin{figure}
  \includegraphics[width=0.467\textwidth]{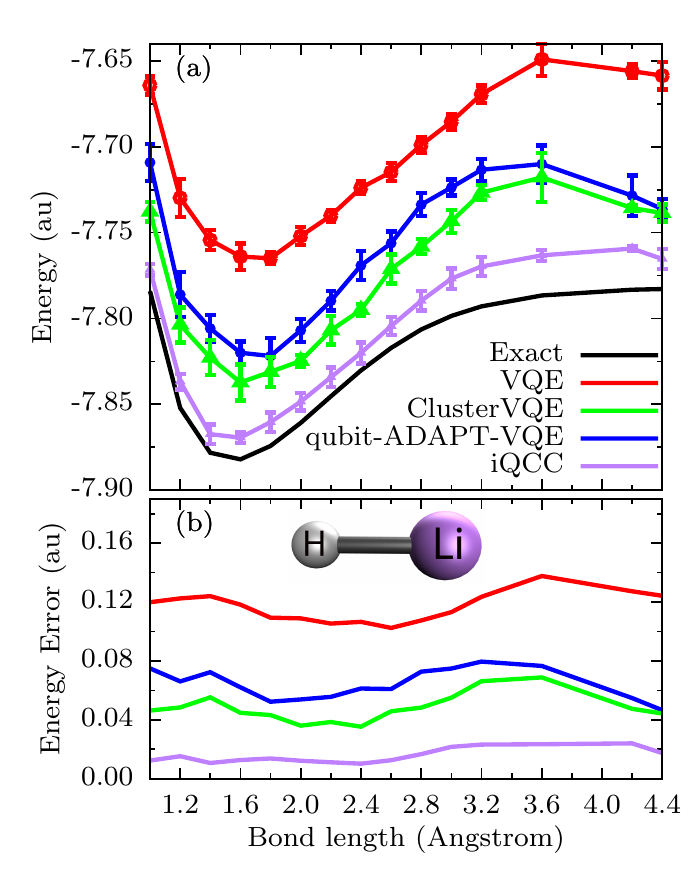}
  \caption{\label{fig_err}  (a) Dissociation curves of LiH and (b) corresponding errors with exact diagonalization (black), VQE (red), qubit-ADAPT-VQE (blue), iQCC (purple), and ClusterVQE (green) methods on a noisy simulator. 
  The custom noise in Qiskit~\citep{Qiskit} was used, where the error rates of single-qubit and two-qubit gates are 0.001 and 0.005, respectively. The means and error bars are calculated from 5 independent simulations. The bottom panel shows the corresponding errors. 
  }
\end{figure}

We next examine the dissociation curve of the LiH molecule in the presence of noise to demonstrate ClusterVQE's resilience to this common feature of all NISQ architectures. 
Figure~\ref{fig_err} shows the dissociation curve of the LiH molecule calculated with different algorithms. The energy values are averaged over 5 runs. As shown by the curves, in the presence of noise, the dissociation curves obtained from the VQE, qubit-ADAPT-VQE, iQCC, and ClusterVQE methods are different from the exact solution. Since the VQE algorithm requires more parameters and the deepest circuit, the noise influence is the largest across the set, showing the strongest deviation from the reference. The qubit-ADAPT-VQE method, in contrast, is able to grow a compact ansatz with fewer parameters and lower circuit depth. The influence of noise is suppressed. On the other hand, the iQCC algorithm uses much shallower circuit depth. As a result, the iQCC is most immune to the noise, and the corresponding dissociation curve is the closest to the reference. Next is ClusterVQE which partially dresses the Hamiltonian and grows the ansatz. The circuit depth of ClusterVQE lies between that of iQCC and qubit-ADAPT-VQE. Consequently, the energy error in the presence of noise is also in between that of the iQCC and qubit-ADAPT-VQE methods. 

We also compare the performance of the four algorithms with different error rates. Figures~\ref{fig_gate_err}(a) and (b) compare the energy errors with respect to different single- and two-qubit error rates, respectively. The single-qubit gates usually have much lower error rates on real quantum devices. For example, the single- and two-qubit gate error rates of the IBM-Yorktown~\cite{IBM2020} device are around $\sim10^{-3}$ and $\sim 10^{-2}$, respectively. Consequently, the error of single-qubit gates have a much smaller influence on the results.  In contrast, the results are significantly affected by the error of two-qubit gates. In general, the error of ClusterVQE lies between that of the qubit-ADAPT-VQE and iQCC methods on the noisy simulator, as expected.

In the previous examples, the MI is calculated based on the exact wavefunction (WF) for proof-of-principle. But, the exact WF is not known before the VQE calculation. However, an approximate calculation of MI provides a sufficient estimate of the correlation between different qubits, as shown in the previous DMRG-type selection of active space~\cite{dmrg_active} and our recent development of the PermVQE algorithm~\cite{permVQE2021}. The approximate MI can be estimated on classical computers by using either MP2 or DMRG (with small bond-dimension) methods. Here, we also demonstrate that ClusterVQE even works with MI calculated on-the-fly. At each iteration, the MI is updated and used for re-clustering the qubits. As shown by the red dots in Figures~\ref{fig_lih} (a) and (c), the performance of ClusterVQE on-the-fly is almost the same as that using the exact MI. The clustering is suboptimal only for the first three iterations
and the dressed Hamiltonian has a slightly larger size as a result. The compositions of each cluster are changing before a stable clustering is found. Nevertheless, these results illustrate that ClusterVQE also works when clustering on-the-fly.

\begin{figure}
  \includegraphics[width=0.5\textwidth]{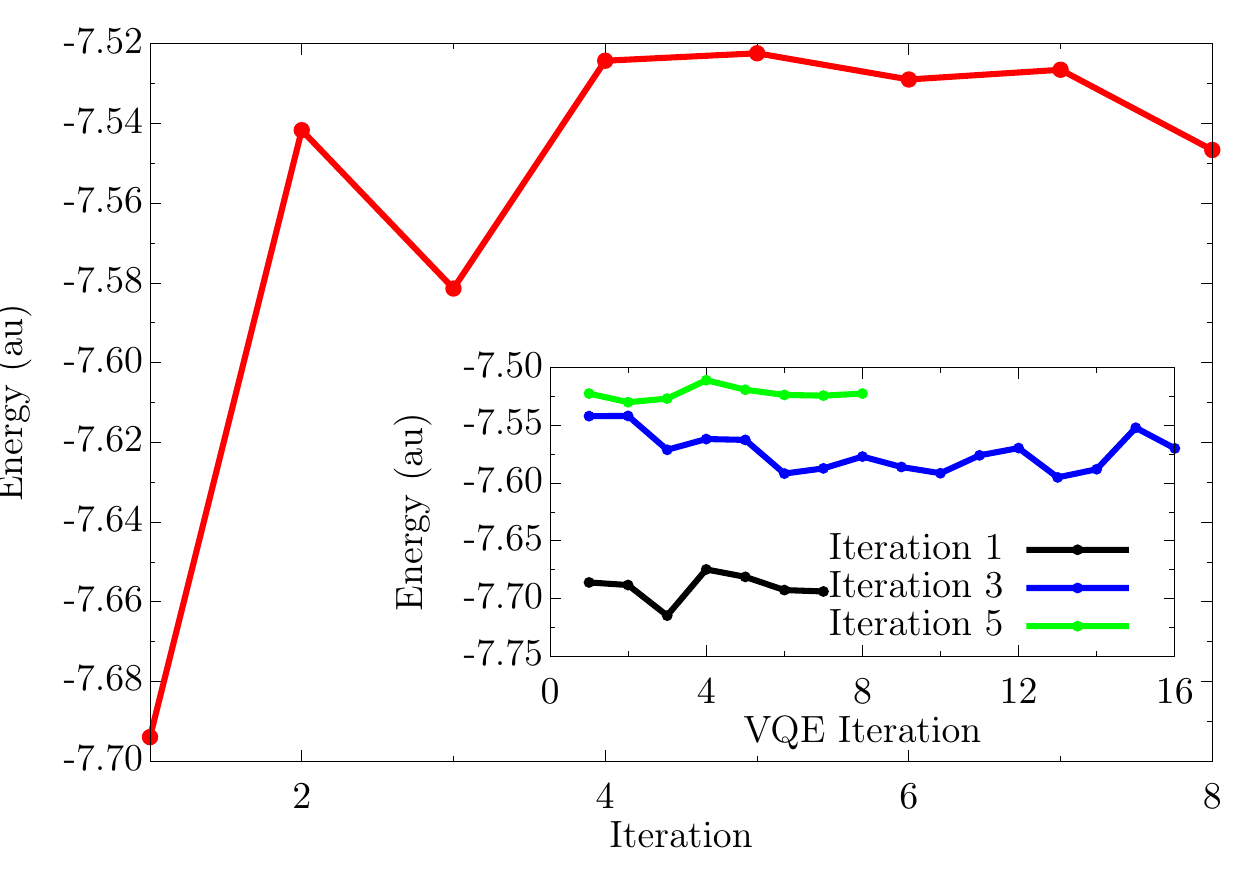}
  \caption{\label{fig_rome} Simulation of the LiH molecule (10 qubits) on the IBMQ-Rome quantum device (5 qubits). The error rate of the CNOT gate is $0.0067\sim 0.026$ and 0.014 in average at runtime. The inset plot shows the VQE optimization for the first, third and fifth ClusterVQE iterations.}
\end{figure}

\subsection{Simulation on a quantum computer}
Finally, we implement ClusterVQE on an actual quantum backend. The IBMQ-Rome device with 5 qubits was chosen to simulate the LiH molecule described by 10 qubits (lowest energy spatial orbital is assumed to be frozen). In order to run on IBMQ-Rome, we set the size of each cluster to be 5. Figure~\ref{fig_rome} shows the energies as a function of ClusterVQE iterations without error mitigation, and the inset plot shows the VQE optimization for the first, third, and fifth ClusterVQE iterations, respectively. Due to the large CNOT error rate, the energies go up with the first a few ClusterVQE iterations without error mitigation. Namely, the number of entanglers encoded on the quantum circuits increases with the number of iterations and introduces more errors. However, the energy error does not increase monotonically with ClusterVQE iterations, suggesting that the ClusterVQE can compete with noise even without error mitigation. 

The ClusterVQE performance on real quantum devices can be further improved with recently developed error mitigation techniques~\cite{LZ_error2021}. 
For proof-of-principle, we first perform ClusterVQE on the ideal simulator to generate the circuits of the final iteration and run the final circuits on both the noisy simulator and real quantum device (IBMQ-Rome) for error mitigation. We perform the mitigation using Clifford data regression (CDR)~\cite{czarnik2020error},~\footnote{CDR training set is constructed using 4 near-Clifford circuits with 1 non-Clifford gate in each of 2 clusters. The training circuits are chosen by replacing non-Clifford gates with the closest Clifford gates.}. For the noisy simulation, we use a noise model obtained by gate set tomography of the IBMQ Ourense device~\cite{cincio2021machine}. The results are listed in Table.~\ref{tab_errs}, which shows that error mitigation reduces the error by several orders of magnitude. Such simulations demonstrate that the ClusterVQE method, with error mitigation, can readily achieve chemical accuracy on real quantum devices. 

%

\begin{table}[!htb]
\caption{\label{tab_errs} Energy errors with and without error mitigation. We show results for different numbers of shots  per  measurement of a  Hamiltonian term.     
}
\begin{tabular}{c|c|c|c}
\hline
\multirow{2}{*}{Backend} & \multirow{2}{*}{Shots} & \multicolumn{2}{c}{Energy error (au)} \\ \cline{3-4} 
&  & Not mitigated  &  Mitigated    \\ \hline
Noisy simulator & $204080$ & 0.225 & $5.6\times 10^{-2}$ \\ \hline
Noisy simulator & $\infty$ & 0.221 & $1.0\times 10^{-5}$ \\ \hline
IBMQ-Rome & 5120 & 0.236 & $1.98\times 10^{-3}$ \\ \hline
\end{tabular}
\end{table}

\section{Conclusion} 
To summarize, we developed a novel ClusterVQE algorithm to reduce quantum circuit complexity, including both the number of qubits and circuit depth. Our algorithm groups qubits into clusters based on maximal correlation between them and 
further takes into account the residual entanglement between clusters in a new "dressed" Hamiltonian. Hence, ClusterVQE naturally implements the benefits of the previously developed PermVQE algorithm, which significantly reduces the number of controlled-NOT gates to connect strongly correlated qubits~\cite{permVQE2021}. The entanglers that couple different clusters are further used to dress the Hamiltonian and remove the entanglement between different clusters. Consequently, the Hamiltonian can be decomposed as a tensor product of each cluster. Each cluster can then be measured separately with a smaller number of qubits and shallower circuit depth. Therefore, ClusterVQE is a particularly promising algorithmic development for quantum chemistry applications, where it is able to reduce the quantum hardware requirements significantly. Moreover, our algorithm has the flexibility of reducing qubits to any number at the expense of additional computational costs on classical computers. Since fewer entanglers are encoded on quantum circuits, ClusterVQE is more robust to noise than qubit-ADAPT-VQE, being one of the most circuit-efficient VQE algorithms, which is verified by our numerical experiments on both the noisy simulator and real quantum platforms. Since the correlation between different clusters is minimized, only a few operators are used to dress the Hamiltonian. Consequently, the Hamiltonian size does not increase significantly compared to that of the iQCC method. 

ClusterVQE also uses a newly proposed gradient measurement method for the VQE without using an ancillary qubit. The analytical gradient can significantly improve the performance of ClusterVQE in noisy environments. The simulation of the LiH molecule on the noisy quantum simulator found that the energy error of the ClusterVQE algorithm is smaller than that of the qubit-ADAPT-VQE method due to the shallower circuit depth. 
A groundbreaking example of the exact error-corrected LiH molecule simulation described by 10 qubits on the 5-qubits IBMQ-Rome quantum computer with ClusterVQE is demonstrated. Obtained results further demonstrate the validity and advantage of the ClusterVQE algorithm. Hence, we believe the algorithm developed in this work can pave the way towards the adaptation of molecular quantum chemistry simulation on NISQ devices. 
It should be mentioned that since the size of the qubit Hamiltonian increases with the number of operators used to dress the Hamiltonian, it increases both computational costs on classical computers and measurements on quantum computers. The recently developed measurement reduction techniques~\cite{RN818,RN826,RN824,RN817,RN757,RN784,RN626,William2019arxiv} are likely a path-forward to further reduce the simulation cost on quantum devices.


Finally, it should be noted that the computational scaling of ClusterVQE is dependent on the number of dressing operators. 
Even though the scaling of computing the dressed Hamiltonian is exponential against the number of dressing operators~\cite{iqcc2020}, the scaling of ClusterVQE against iterations is not exponential in general because not every entangler is used to dress the Hamiltonian. 
For the best cases, only a limited number of entanglers is used to dress the Hamiltonian (LiH, for example). Such a situation can be intuitively expected in molecular multimers, where the intermolecular coupling is usually smaller than the intramolecular correlation. For the worse cases, usually in strongly correlated systems in which qubit clustering is not optimal, the scaling of ClusterVQE may be exponential if a larger number of entanglers are used to dress the Hamiltonian. However, even in strongly correlated molecules, the scaling of different geometries can be different. For example, the ClusterVQE simulation of 14-qubits $N_2$ at equilibrium ends up with a large number of Pauli words (Figure~\ref{fign2}(b)). However, the simulation of $N_2$ with stretched bond has a much better scaling (Figure~\ref{fign2}(d)) even though the stretched configuration has a stronger correlation. That being said, the bottleneck of our approach relies on whether the inter-cluster MI can be efficiently minimized. Hence, ClusterVQE does not introduce exponential scaling for general cases.

\begin{acknowledgments}
Research presented in this article was supported by the Laboratory Directed Research and Development (LDRD) program of Los Alamos National Laboratory (LANL) under project number 20200056DR. LANL is operated by Triad National Security, LLC, for the National Nuclear Security Administration of U.S. Department of Energy (contract no. 89233218CNA000001). We thank LANL Institutional Computing (IC) program for access to HPC resources. 
PJC and LC were partially supported by the U.S. Department of Energy (DOE), Office of Science, Office of Advanced Scientific Computing Research, under the Accelerated Research in Quantum Computing (ARQC) program. Quantum resources from IBMQ Hub are acknowledged. This research was also supported by the U.S. Department of Energy (DOE) National Nuclear Security Administration (NNSA) Advanced Simulation and Computing (ASC) program at LANL. We acknowledge the ASC program at LANL for use of their Ising D-Wave 2000Q quantum computing resource.

\end{acknowledgments}

\begin{center}
{\Large \textbf{Appendix}}
\end{center}

\setcounter{section}{0}
\setcounter{subsection}{0}
\setcounter{subsubsection}{0}
\setcounter{equation}{0}
\setcounter{figure}{0}

\renewcommand{\theequation}{S\arabic{equation}}
\renewcommand\thefigure{S\arabic{figure}}

\section{Clustering Methods Used}\label{appendix1}
Three clustering methods were explored for this effort. The details of each are described below.

\subsection{Metis Graph Partitioning on CPU}\label{app_metis}
The MI matrix can be viewed as a graph. This allows for the use of classical graph partitioning methods from the Metis library~\cite{metis}. The \textit{part$\_$graph} function in the Metis library was used and the partition weights were also optimized to minimize the cut. 

\subsection{MI-based clustering on Quantum Annealer QPU}\label{app_gauge}
The MI between qubits can be interpreted as a degree of correlation. Note that when two random variables are fully correlated, their joint entropy is minimal, and the MI is maximal. Moreover, MI can be mapped to what others have coined as the ``generalized correlation coefficient", where all higher order correlations are taking into account \cite{Lange2008-aq,Lange2005-if,Rivalta2012-en}. This generalized correlation coefficient is computed as: 
\begin{equation}
    r^{MI}_{ij} = [ 1 - \mathrm{e}^{-2I_{ij}/d}]^{1/2}
\label{corr}
\end{equation}
where $d$ is the dimensionality of the problem, and $I_{ij}$ are is the Mutual Information matrix elements. Note that, when $I_{ij}$ goes to infinity, $r^{MI}_{ij}$ approaches 1; whereas if $I_{ij}$ tends to zero, $r^{MI}_{ij}$ tends to 0.
Taking into account Equation~\ref{corr}, we see that, by maximizing MI for a constrained set of qubits, we will be maximizing the correlation among that same set. 
Moreover, maximizing the MI restricted to a certain number of qubits, is somehow related to getting the ``most central hub," according to node centrality ordering. Albeit for some exceptions, we have verified the latter using eigenvector centrality metric.  

Maximizing the MI restricted to a certain number of qubits can be formulated as a quadratic unconstrained binary optimization (QUBO) problem. This formulation can, in turn, benefit from available alternative computational paradigms tailored to solve these kinds of problems, e.g. Quantum, and Digital annealers. The cost function to be maximized can be written as: 
\begin{equation}
    C = \textbf{x} (I - \lambda R) \textbf{x}^t  
\label{corr}
\end{equation}
where the vector $\textbf{x}$ is a bit-string containing the information of the selected qubits. This is, if the i-th qubit is selected $x_i = 1$, and 0 otherwise. %
Matrix $I$ contains all the MI elements $I_{ij}$ as defined by Eq.~\ref{eq_mutual}, and $R$ is the matrix representation of an operator that ``restricts" the total number of selected qubits \cite{Rogers2020-fc}.
A penalty constant $\lambda$ is used to tune the strength of the constraint. 
Since the annealing methods and devices commonly minimize cost function, the problem is rather solved by minimizing $-C$ leading to a QUBO matrix equal to $Q = -I + \lambda R$

Matrix R, is defined as follows.
\[
        R_{ij} = 
        \begin{cases}
            1 &\quad\text{if}\quad i \neq j \\
            1 - \alpha \times f \times S &\quad\text{if} \quad i = j 
        \end{cases}
\]
Here, $f$ is the target fraction of qubits to be selected; $S$ is the total size of the qubits pool; and $\alpha$ is a constant that depends on the optimizer (1 or 2 depending if only the upper triangle vs the full QUBO matrix is used to compute the energy).
For all the calculations where this clustering method was used we have set $f = 0.5$, and $\lambda = 2.0$.

An explanation of why the $R$ matrix restricts the set is described next.
Consider the case where a general vector $\mathbf{x}$ of length $S = N$ is such that $\sum_k x_k= K$.
Now let us compute the energy term of that vector given by matrix $R$. This is  $E_R(\mathbf{x}, \mathrm{such\, that \,} ||x||_1 = K) = \mathbf{x}^t R \mathbf{x}$.
Then, the off diagonal elements of $R$ will contribute with $K(K-1)$. If we want  $f = M/N$, then, the diagonal elements will contribute with $K - \alpha MK$. By summing the two contributions, we have that $E_R(K) = K^2 - \alpha MK = K(K - \alpha M)$. 
This means that the function will have two minima, the trivial one at K=0, and another at $K = \alpha M$ ($\alpha M$ bits equal to 1). 

\subsection{Quantum Community Detection on Quantum Annealer QPU}\label{qcomm_detection}
Classical community detection as graph clustering maximizes modularity to split the MI matrix into clusters that minimize correlation or connectivity between clusters~\cite{Newman2006}. In quantum community detection (QCD), the problem is represented as a QUBO that can be run on a D-Wave quantum annealer as an NP-complete combinatorial optimization problem.

The symmetric MI matrix is viewed as a graph. The adjacency matrix $A$ is calculated as follows.
\[
  A_{ij} = 
  \begin{cases} 
    0 & i = j \\
    |{MI}_{ij}| & i \neq j
  \end{cases}
\]

The modularity matrix $B$ is calculated from the adjacency matrix $A$ and degree of each node (number of connected nodes) $d_i$.
\begin{equation}
    B_{ij} =  A_{ij} - \frac{d_id_j}{2m} =  A_{ij} - \frac{d_id_j}{\sum_{i}d_{i}}
    \label{eq:ModM}
\end{equation}

While QCD can cluster a graph into 2 or more clusters~\cite{Ushijima-Mwesigwa2017,Negre2020,sue_qcd2021}, in this work we are interested in splitting only into 2 clusters. The modularity matrix in QUBO form is run on the D-Wave 2000Q quantum annealer: 
\begin{equation}
    Q(x) = max(x^TBx)
    \label{eq:mod_obj}
\end{equation}

The resulting $x$ is a string of zeros and ones that maps each of the MI nodes (representing qubits) to one of two clusters that can be used in the ClusterVQE algorithm.

\begin{figure}[!h]
  \includegraphics[width=0.5\textwidth]{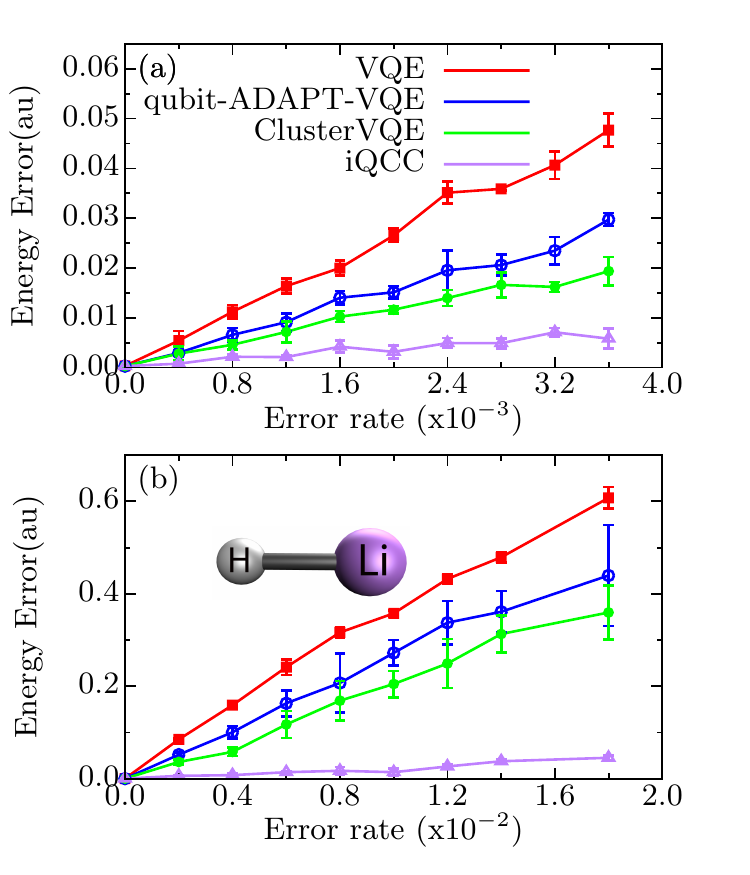}
  \caption{\label{fig_gate_err} The error in energy ($\Delta E$) as a function of the error rates of single (a) and two-qubit (b) gates, respectively. The two and single-qubit gate error in a and b are fixed at 0.01 and 0.001, respectively. The single-qubit gate on real quantum devices is usually one order of magnitude smaller than that of two-qubit gates. Hence, single qubit error has a smaller impact on the results.}
\end{figure}

\section{Introduction to UCCSD ansatz and Qubit-UCCSD}
\label{appuccsd}
Within the UCCSD ansatz, the correlation beyond the Hartree-Fock reference state is introduced via the single and double excitation (and de-excitation) operators:
$\ket{\Phi(\theta)}=e^{\hT(\theta)-\hT^\dag(\theta)}\ket{\Phi_r}$,
where $\ket{\Phi_r}$ is the reference state being the HF state in this work. $\hT(\theta)=\hT_1(\theta_1)+\hT_2(\theta_2)$ is the excitation operator with $\hT_1(\theta_1)$ and $\hT_2(\theta_2)$ being the parameterized single and double excitation operators, respectively, 
\begin{eqnarray}
 \hT_1(\theta) = && \sum_{ia}\theta^a_i \hc^\dag_a\hc_i \nonumber\\
 \hT_2(\theta) = && \sum_{ij,ab}\theta^{ab}_{ij}\hc^\dag_a\hc^\dag_b \hc_i\hc_j.
\end{eqnarray}
For simplicity, indices $i, j, \cdots$ are used to represent the occupied molecular orbitals (MOs), $a, b, \cdots$ represent the unoccupied MOs.  Because the NISQ devices only have gates acting on a few qubits at a time, the UCCSD operator must be broken up into a sequence of single operators (either single or double excitation) via the Trotter expansion of an exponent operator~\cite{SUZUKI1990319}, i.e.,  $e^{A+B}=\lim_{n\rightarrow\infty}\left(e^{A/n }e^{B/n}\right)^n$.  Even though the truncated Trotter expansion is an approximation to the UCCSD ansatz, the variational procedure is able to absorb the error according to recent report~\cite{PhysRevX.6.031007}. Consequently, a finite order of Trotter expansion is able to reproduce the FCI energies due to the fact that the many-body interaction can be decomposed as the tensor product of one- and two-body interactions~\cite{RN170}. In this way, the exact FCI state can be written as the tensor product of single and double excitation operators, $\ket{\Phi(\theta)}=\prod^D_{k=1} e^{\ot_k(\theta)}\ket{\Phi_r}$, where $\ot_k(\theta)$ represents either single or double excitation operators and their anti-Hermitian forms ($\ot^\dag_k=-\ot_k$). Consequently, the GS energy is readily obtained by minimizing the parameterized ansatz, i.e.,
\begin{eqnarray}\label{eq-minegs}
E_{GS}=&&\min_{\theta}\bra{\Phi(\theta)}\hH\ket{\Phi(\theta)}\nonumber\\
=&&\min_{\theta}\bra{\Phi_r}\prod^1_{k=D} e^{-\ot_k(\theta_k)}\hH\prod^D_{k=1} e^{\ot_k(\theta_k)}\ket{\Phi_r}
\end{eqnarray} 
In the UCCSD ansatz, the SD sequences are directly generated from the Trotter expansion of the $\hT(\theta)-\hT^\dag(\theta)$ operator. In the recently developed ADAPT-VQE algorithm~\cite{RN170}, the sequence of SD operators in the ansatz are chosen adaptively by measuring their contribution to the GS energy weighted by the energy gradients with respect to the parameters. With the adaptive method, a more compact sequence of the unitary operators is generated and a smaller number of parameters is optimized at the cost of more VQE simulations. Because the UCCSD ansatz is not hardware-efficient, a qubit version of the UCCSD ansatz was developed recently in which the ansatz is broken up into a series of Pauli words and a small portion of the Pauli words is chosen to adaptively grow the ansatz. Eq.~\ref{eq:derivative} indicates the product $\hP_{kj}=[\hP_k,\hP_j]$, that only contains the $\hZ$ operators, has non-vanishing derivatives because $\bra{\Phi_r}\hat{\sigma}_i\ket{\Phi_r}=0$ if $\hat{\sigma}_i=\hX,\hY$. Therefore, the Pauli word $\hP_j$ that contains $\hZ$ has no contribution to the derivative. And two Pauli words with the the same flip indices (indices of $\hX/\hY$) have the same contribution. 
Hence, after the fermion-qubit mapping of the UCCSD operators, we only include those Pauli words with unique flip indices. Thus, only one of these operators is included in the unitary operators $\{\hU_j\}$. In this way, the number of operators and circuit depth can be reduced. 

\section{Analytical gradient in VQE}\label{appgradient}
The energy gradient is used in the parameter optimization on the classical computers. The gradient can be obtained by the finite-difference method or analytical gradients. According to our numerical experiment on either a noisy simulator or real quantum devices, the analytical gradient can significantly improve the performance of VQE in the presence of noise. In this work, the analytical gradient is obtained by the measurements on quantum circuits. The gradient of a parameter is,
\begin{equation}
\frac{\partial E}{\partial \theta_i}=2\text{Im}\left\{\bra{\Phi_r}\hU^\dag \hH\frac{\partial \hU}{\partial\theta_i}\ket{\Phi_r}\right\}.
\end{equation}
Suppose there are $M$ operators in the ansatz, i.e, $\hU=\prod^M_{k=1}\hU_k$, the derivative of $\hU$ with respect to $\theta_i$ reads
\begin{eqnarray}
\frac{\partial\hU}{\partial\theta_i}=&&i\prod^M_{k=i+1}\hU_k \hP_i\prod^{i}_{k=1}\hU_k
\nonumber\\
=&&i\left(\prod^M_{k=i+1}\hU_k \hP_i \prod^{i+1}_{k=M}\hU^\dag_k\right)\prod^{M}_{k=1}\hU_k
\equiv i\hP^{\text{eff}}_i\hU,
\end{eqnarray}
where $\hP^{\text{eff}}_i$ is the dressed operator. Consequently, the energy gradient can be obtained by
\begin{equation}
\frac{\partial E}{\partial \theta_i}=2\text{Im}\left\{\bra{\Phi_r}\hU^\dag \hH\hP^{\text{eff}}_i\hU\ket{\Phi_r}\right\}.
\end{equation}
In this way, calculation of the dressed operator $\hP^{\text{eff}}_i$ on classical computers is required, while no ancillary qubit on the quantum circuit is needed.


\bibliography{ref}

\begin{thebibliography}{85}%
\makeatletter
\providecommand \@ifxundefined [1]{%
 \@ifx{#1\undefined}
}%
\providecommand \@ifnum [1]{%
 \ifnum #1\expandafter \@firstoftwo
 \else \expandafter \@secondoftwo
 \fi
}%
\providecommand \@ifx [1]{%
 \ifx #1\expandafter \@firstoftwo
 \else \expandafter \@secondoftwo
 \fi
}%
\providecommand \natexlab [1]{#1}%
\providecommand \enquote  [1]{``#1''}%
\providecommand \bibnamefont  [1]{#1}%
\providecommand \bibfnamefont [1]{#1}%
\providecommand \citenamefont [1]{#1}%
\providecommand \href@noop [0]{\@secondoftwo}%
\providecommand \href [0]{\begingroup \@sanitize@url \@href}%
\providecommand \@href[1]{\@@startlink{#1}\@@href}%
\providecommand \@@href[1]{\endgroup#1\@@endlink}%
\providecommand \@sanitize@url [0]{\catcode `\\12\catcode `\$12\catcode
  `\&12\catcode `\#12\catcode `\^12\catcode `\_12\catcode `\%12\relax}%
\providecommand \@@startlink[1]{}%
\providecommand \@@endlink[0]{}%
\providecommand \url  [0]{\begingroup\@sanitize@url \@url }%
\providecommand \@url [1]{\endgroup\@href {#1}{\urlprefix }}%
\providecommand \urlprefix  [0]{URL }%
\providecommand \Eprint [0]{\href }%
\providecommand \doibase [0]{http://dx.doi.org/}%
\providecommand \selectlanguage [0]{\@gobble}%
\providecommand \bibinfo  [0]{\@secondoftwo}%
\providecommand \bibfield  [0]{\@secondoftwo}%
\providecommand \translation [1]{[#1]}%
\providecommand \BibitemOpen [0]{}%
\providecommand \bibitemStop [0]{}%
\providecommand \bibitemNoStop [0]{.\EOS\space}%
\providecommand \EOS [0]{\spacefactor3000\relax}%
\providecommand \BibitemShut  [1]{\csname bibitem#1\endcsname}%
\let\auto@bib@innerbib\@empty
\bibitem [{\citenamefont {{Gan}}\ and\ \citenamefont
  {{Harrison}}(2005)}]{1559974}%
  \BibitemOpen
  \bibfield  {author} {\bibinfo {author} {\bibfnamefont {Z.}~\bibnamefont
  {{Gan}}}\ and\ \bibinfo {author} {\bibfnamefont {R.~J.}\ \bibnamefont
  {{Harrison}}},\ }in\ \href@noop {} {\emph {\bibinfo {booktitle} {SC '05:
  Proceedings of the 2005 ACM/IEEE Conference on Supercomputing}}}\ (\bibinfo
  {year} {2005})\ pp.\ \bibinfo {pages} {22--22}\BibitemShut {NoStop}%
\bibitem [{\citenamefont {Cao}\ \emph {et~al.}(2019)\citenamefont {Cao},
  \citenamefont {Romero}, \citenamefont {Olson}, \citenamefont {Degroote},
  \citenamefont {Johnson}, \citenamefont {Kieferov{\'a}}, \citenamefont
  {Kivlichan}, \citenamefont {Menke}, \citenamefont {Peropadre}, \citenamefont
  {Sawaya} \emph {et~al.}}]{cao2018quantum}%
  \BibitemOpen
  \bibfield  {author} {\bibinfo {author} {\bibfnamefont {Y.}~\bibnamefont
  {Cao}}, \bibinfo {author} {\bibfnamefont {J.}~\bibnamefont {Romero}},
  \bibinfo {author} {\bibfnamefont {J.~P.}\ \bibnamefont {Olson}}, \bibinfo
  {author} {\bibfnamefont {M.}~\bibnamefont {Degroote}}, \bibinfo {author}
  {\bibfnamefont {P.~D.}\ \bibnamefont {Johnson}}, \bibinfo {author}
  {\bibfnamefont {M.}~\bibnamefont {Kieferov{\'a}}}, \bibinfo {author}
  {\bibfnamefont {I.~D.}\ \bibnamefont {Kivlichan}}, \bibinfo {author}
  {\bibfnamefont {T.}~\bibnamefont {Menke}}, \bibinfo {author} {\bibfnamefont
  {B.}~\bibnamefont {Peropadre}}, \bibinfo {author} {\bibfnamefont {N.~P.}\
  \bibnamefont {Sawaya}},  \emph {et~al.},\ }\href
  {https://pubs.acs.org/doi/10.1021/acs.chemrev.8b00803} {\bibfield  {journal}
  {\bibinfo  {journal} {Chem. Rev.}\ }\textbf {\bibinfo {volume} {119}},\
  \bibinfo {pages} {10856} (\bibinfo {year} {2019})}\BibitemShut {NoStop}%
\bibitem [{\citenamefont {McArdle}\ \emph {et~al.}(2020)\citenamefont
  {McArdle}, \citenamefont {Endo}, \citenamefont {Aspuru-Guzik}, \citenamefont
  {Benjamin},\ and\ \citenamefont {Yuan}}]{mcardle2020quantum}%
  \BibitemOpen
  \bibfield  {author} {\bibinfo {author} {\bibfnamefont {S.}~\bibnamefont
  {McArdle}}, \bibinfo {author} {\bibfnamefont {S.}~\bibnamefont {Endo}},
  \bibinfo {author} {\bibfnamefont {A.}~\bibnamefont {Aspuru-Guzik}}, \bibinfo
  {author} {\bibfnamefont {S.~C.}\ \bibnamefont {Benjamin}}, \ and\ \bibinfo
  {author} {\bibfnamefont {X.}~\bibnamefont {Yuan}},\ }\href
  {https://journals.aps.org/rmp/abstract/10.1103/RevModPhys.92.015003}
  {\bibfield  {journal} {\bibinfo  {journal} {Rev. Mod. Phys.}\ }\textbf
  {\bibinfo {volume} {92}},\ \bibinfo {pages} {015003} (\bibinfo {year}
  {2020})}\BibitemShut {NoStop}%
\bibitem [{\citenamefont {Kitaev}(1995)}]{kitaev1995quantum}%
  \BibitemOpen
  \bibfield  {author} {\bibinfo {author} {\bibfnamefont {A.~Y.}\ \bibnamefont
  {Kitaev}},\ }\href@noop {} {\enquote {\bibinfo {title} {Quantum measurements
  and the abelian stabilizer problem},}\ } (\bibinfo {year} {1995}),\ \Eprint
  {http://arxiv.org/abs/quant-ph/9511026} {arXiv:quant-ph/9511026 [quant-ph]}
  \BibitemShut {NoStop}%
\bibitem [{\citenamefont {Aspuru-Guzik}\ \emph {et~al.}(2005)\citenamefont
  {Aspuru-Guzik}, \citenamefont {Dutoi}, \citenamefont {Love},\ and\
  \citenamefont {Head-Gordon}}]{Aspuru-Guzik1704}%
  \BibitemOpen
  \bibfield  {author} {\bibinfo {author} {\bibfnamefont {A.}~\bibnamefont
  {Aspuru-Guzik}}, \bibinfo {author} {\bibfnamefont {A.~D.}\ \bibnamefont
  {Dutoi}}, \bibinfo {author} {\bibfnamefont {P.~J.}\ \bibnamefont {Love}}, \
  and\ \bibinfo {author} {\bibfnamefont {M.}~\bibnamefont {Head-Gordon}},\
  }\href {\doibase 10.1126/science.1113479} {\bibfield  {journal} {\bibinfo
  {journal} {Science}\ }\textbf {\bibinfo {volume} {309}},\ \bibinfo {pages}
  {1704} (\bibinfo {year} {2005})}\BibitemShut {NoStop}%
\bibitem [{\citenamefont {Preskill}(2018)}]{preskil2018}%
  \BibitemOpen
  \bibfield  {author} {\bibinfo {author} {\bibfnamefont {J.}~\bibnamefont
  {Preskill}},\ }\href@noop {} {\bibfield  {journal} {\bibinfo  {journal}
  {Quantum}\ }\textbf {\bibinfo {volume} {2}},\ \bibinfo {pages} {79} (\bibinfo
  {year} {2018})}\BibitemShut {NoStop}%
\bibitem [{\citenamefont {Wei}\ \emph {et~al.}(2020)\citenamefont {Wei},
  \citenamefont {Li},\ and\ \citenamefont {Long}}]{RN129}%
  \BibitemOpen
  \bibfield  {author} {\bibinfo {author} {\bibfnamefont {S.}~\bibnamefont
  {Wei}}, \bibinfo {author} {\bibfnamefont {H.}~\bibnamefont {Li}}, \ and\
  \bibinfo {author} {\bibfnamefont {G.}~\bibnamefont {Long}},\ }\href {\doibase
  10.34133/2020/1486935} {\bibfield  {journal} {\bibinfo  {journal} {Research}\
  }\textbf {\bibinfo {volume} {2020}},\ \bibinfo {pages} {1486935} (\bibinfo
  {year} {2020})}\BibitemShut {NoStop}%
\bibitem [{\citenamefont {Moll}\ \emph {et~al.}(2018)\citenamefont {Moll},
  \citenamefont {Barkoutsos}, \citenamefont {Bishop}, \citenamefont {Chow},
  \citenamefont {Cross}, \citenamefont {Egger}, \citenamefont {Filipp},
  \citenamefont {Fuhrer}, \citenamefont {Gambetta}, \citenamefont {Ganzhorn},
  \citenamefont {Kandala}, \citenamefont {Mezzacapo}, \citenamefont {Müller},
  \citenamefont {Riess}, \citenamefont {Salis}, \citenamefont {Smolin},
  \citenamefont {Tavernelli},\ and\ \citenamefont {Temme}}]{Moll_2018}%
  \BibitemOpen
  \bibfield  {author} {\bibinfo {author} {\bibfnamefont {N.}~\bibnamefont
  {Moll}}, \bibinfo {author} {\bibfnamefont {P.}~\bibnamefont {Barkoutsos}},
  \bibinfo {author} {\bibfnamefont {L.~S.}\ \bibnamefont {Bishop}}, \bibinfo
  {author} {\bibfnamefont {J.~M.}\ \bibnamefont {Chow}}, \bibinfo {author}
  {\bibfnamefont {A.}~\bibnamefont {Cross}}, \bibinfo {author} {\bibfnamefont
  {D.~J.}\ \bibnamefont {Egger}}, \bibinfo {author} {\bibfnamefont
  {S.}~\bibnamefont {Filipp}}, \bibinfo {author} {\bibfnamefont
  {A.}~\bibnamefont {Fuhrer}}, \bibinfo {author} {\bibfnamefont {J.~M.}\
  \bibnamefont {Gambetta}}, \bibinfo {author} {\bibfnamefont {M.}~\bibnamefont
  {Ganzhorn}}, \bibinfo {author} {\bibfnamefont {A.}~\bibnamefont {Kandala}},
  \bibinfo {author} {\bibfnamefont {A.}~\bibnamefont {Mezzacapo}}, \bibinfo
  {author} {\bibfnamefont {P.}~\bibnamefont {Müller}}, \bibinfo {author}
  {\bibfnamefont {W.}~\bibnamefont {Riess}}, \bibinfo {author} {\bibfnamefont
  {G.}~\bibnamefont {Salis}}, \bibinfo {author} {\bibfnamefont
  {J.}~\bibnamefont {Smolin}}, \bibinfo {author} {\bibfnamefont
  {I.}~\bibnamefont {Tavernelli}}, \ and\ \bibinfo {author} {\bibfnamefont
  {K.}~\bibnamefont {Temme}},\ }\href {\doibase 10.1088/2058-9565/aab822}
  {\bibfield  {journal} {\bibinfo  {journal} {Quantum Sci. Technol.}\ }\textbf
  {\bibinfo {volume} {3}},\ \bibinfo {pages} {030503} (\bibinfo {year}
  {2018})}\BibitemShut {NoStop}%
\bibitem [{\citenamefont {Cerezo}\ \emph {et~al.}(2020)\citenamefont {Cerezo},
  \citenamefont {Arrasmith}, \citenamefont {Babbush}, \citenamefont {Benjamin},
  \citenamefont {Endo}, \citenamefont {Fujii}, \citenamefont {McClean},
  \citenamefont {Mitarai}, \citenamefont {Yuan}, \citenamefont {Cincio},\ and\
  \citenamefont {Coles}}]{cerezo2020variational}%
  \BibitemOpen
  \bibfield  {author} {\bibinfo {author} {\bibfnamefont {M.}~\bibnamefont
  {Cerezo}}, \bibinfo {author} {\bibfnamefont {A.}~\bibnamefont {Arrasmith}},
  \bibinfo {author} {\bibfnamefont {R.}~\bibnamefont {Babbush}}, \bibinfo
  {author} {\bibfnamefont {S.~C.}\ \bibnamefont {Benjamin}}, \bibinfo {author}
  {\bibfnamefont {S.}~\bibnamefont {Endo}}, \bibinfo {author} {\bibfnamefont
  {K.}~\bibnamefont {Fujii}}, \bibinfo {author} {\bibfnamefont {J.~R.}\
  \bibnamefont {McClean}}, \bibinfo {author} {\bibfnamefont {K.}~\bibnamefont
  {Mitarai}}, \bibinfo {author} {\bibfnamefont {X.}~\bibnamefont {Yuan}},
  \bibinfo {author} {\bibfnamefont {L.}~\bibnamefont {Cincio}}, \ and\ \bibinfo
  {author} {\bibfnamefont {P.~J.}\ \bibnamefont {Coles}},\ }\href@noop {}
  {\enquote {\bibinfo {title} {Variational quantum algorithms},}\ } (\bibinfo
  {year} {2020}),\ \Eprint {http://arxiv.org/abs/2012.09265} {arXiv:2012.09265
  [quant-ph]} \BibitemShut {NoStop}%
\bibitem [{\citenamefont {Bharti}\ \emph {et~al.}(2021)\citenamefont {Bharti},
  \citenamefont {Cervera-Lierta}, \citenamefont {Kyaw}, \citenamefont {Haug},
  \citenamefont {Alperin-Lea}, \citenamefont {Anand}, \citenamefont {Degroote},
  \citenamefont {Heimonen}, \citenamefont {Kottmann}, \citenamefont {Menke},
  \citenamefont {Mok}, \citenamefont {Sim}, \citenamefont {Kwek},\ and\
  \citenamefont {Aspuru-Guzik}}]{bharti2021noisy}%
  \BibitemOpen
  \bibfield  {author} {\bibinfo {author} {\bibfnamefont {K.}~\bibnamefont
  {Bharti}}, \bibinfo {author} {\bibfnamefont {A.}~\bibnamefont
  {Cervera-Lierta}}, \bibinfo {author} {\bibfnamefont {T.~H.}\ \bibnamefont
  {Kyaw}}, \bibinfo {author} {\bibfnamefont {T.}~\bibnamefont {Haug}}, \bibinfo
  {author} {\bibfnamefont {S.}~\bibnamefont {Alperin-Lea}}, \bibinfo {author}
  {\bibfnamefont {A.}~\bibnamefont {Anand}}, \bibinfo {author} {\bibfnamefont
  {M.}~\bibnamefont {Degroote}}, \bibinfo {author} {\bibfnamefont
  {H.}~\bibnamefont {Heimonen}}, \bibinfo {author} {\bibfnamefont {J.~S.}\
  \bibnamefont {Kottmann}}, \bibinfo {author} {\bibfnamefont {T.}~\bibnamefont
  {Menke}}, \bibinfo {author} {\bibfnamefont {W.-K.}\ \bibnamefont {Mok}},
  \bibinfo {author} {\bibfnamefont {S.}~\bibnamefont {Sim}}, \bibinfo {author}
  {\bibfnamefont {L.-C.}\ \bibnamefont {Kwek}}, \ and\ \bibinfo {author}
  {\bibfnamefont {A.}~\bibnamefont {Aspuru-Guzik}},\ }\href@noop {} {\enquote
  {\bibinfo {title} {Noisy intermediate-scale quantum (nisq) algorithms},}\ }
  (\bibinfo {year} {2021}),\ \Eprint {http://arxiv.org/abs/2101.08448}
  {arXiv:2101.08448 [quant-ph]} \BibitemShut {NoStop}%
\bibitem [{\citenamefont {Peruzzo}\ \emph {et~al.}(2014)\citenamefont
  {Peruzzo}, \citenamefont {McClean}, \citenamefont {Shadbolt}, \citenamefont
  {Yung}, \citenamefont {Zhou}, \citenamefont {Love}, \citenamefont
  {Aspuru-Guzik},\ and\ \citenamefont {O'Brien}}]{RN155}%
  \BibitemOpen
  \bibfield  {author} {\bibinfo {author} {\bibfnamefont {A.}~\bibnamefont
  {Peruzzo}}, \bibinfo {author} {\bibfnamefont {J.}~\bibnamefont {McClean}},
  \bibinfo {author} {\bibfnamefont {P.}~\bibnamefont {Shadbolt}}, \bibinfo
  {author} {\bibfnamefont {M.~H.}\ \bibnamefont {Yung}}, \bibinfo {author}
  {\bibfnamefont {X.~Q.}\ \bibnamefont {Zhou}}, \bibinfo {author}
  {\bibfnamefont {P.~J.}\ \bibnamefont {Love}}, \bibinfo {author}
  {\bibfnamefont {A.}~\bibnamefont {Aspuru-Guzik}}, \ and\ \bibinfo {author}
  {\bibfnamefont {J.~L.}\ \bibnamefont {O'Brien}},\ }\href {\doibase
  10.1038/ncomms5213} {\bibfield  {journal} {\bibinfo  {journal} {Nat.
  Commun.}\ }\textbf {\bibinfo {volume} {5}},\ \bibinfo {pages} {4213}
  (\bibinfo {year} {2014})}\BibitemShut {NoStop}%
\bibitem [{\citenamefont {McClean}\ \emph {et~al.}(2016)\citenamefont
  {McClean}, \citenamefont {Romero}, \citenamefont {Babbush},\ and\
  \citenamefont {Aspuru-Guzik}}]{McClean_2016}%
  \BibitemOpen
  \bibfield  {author} {\bibinfo {author} {\bibfnamefont {J.~R.}\ \bibnamefont
  {McClean}}, \bibinfo {author} {\bibfnamefont {J.}~\bibnamefont {Romero}},
  \bibinfo {author} {\bibfnamefont {R.}~\bibnamefont {Babbush}}, \ and\
  \bibinfo {author} {\bibfnamefont {A.}~\bibnamefont {Aspuru-Guzik}},\ }\href
  {\doibase 10.1088/1367-2630/18/2/023023} {\bibfield  {journal} {\bibinfo
  {journal} {New J. Phys.}\ }\textbf {\bibinfo {volume} {18}},\ \bibinfo
  {pages} {023023} (\bibinfo {year} {2016})}\BibitemShut {NoStop}%
\bibitem [{\citenamefont {Li}\ \emph {et~al.}(2019{\natexlab{a}})\citenamefont
  {Li}, \citenamefont {Hu}, \citenamefont {Zhang}, \citenamefont {Song},\ and\
  \citenamefont {Yung}}]{adts.201800182}%
  \BibitemOpen
  \bibfield  {author} {\bibinfo {author} {\bibfnamefont {Y.}~\bibnamefont
  {Li}}, \bibinfo {author} {\bibfnamefont {J.}~\bibnamefont {Hu}}, \bibinfo
  {author} {\bibfnamefont {X.-M.}\ \bibnamefont {Zhang}}, \bibinfo {author}
  {\bibfnamefont {Z.}~\bibnamefont {Song}}, \ and\ \bibinfo {author}
  {\bibfnamefont {M.-H.}\ \bibnamefont {Yung}},\ }\href {\doibase
  https://doi.org/10.1002/adts.201800182} {\bibfield  {journal} {\bibinfo
  {journal} {Adv. Theory Simul.}\ }\textbf {\bibinfo {volume} {2}},\ \bibinfo
  {pages} {1800182} (\bibinfo {year} {2019}{\natexlab{a}})}\BibitemShut
  {NoStop}%
\bibitem [{\citenamefont {Fedorov}\ \emph {et~al.}(2021)\citenamefont
  {Fedorov}, \citenamefont {Peng}, \citenamefont {Govind},\ and\ \citenamefont
  {Alexeev}}]{fedorov2021vqe}%
  \BibitemOpen
  \bibfield  {author} {\bibinfo {author} {\bibfnamefont {D.~A.}\ \bibnamefont
  {Fedorov}}, \bibinfo {author} {\bibfnamefont {B.}~\bibnamefont {Peng}},
  \bibinfo {author} {\bibfnamefont {N.}~\bibnamefont {Govind}}, \ and\ \bibinfo
  {author} {\bibfnamefont {Y.}~\bibnamefont {Alexeev}},\ }\href@noop {}
  {\enquote {\bibinfo {title} {Vqe method: A short survey and recent
  developments},}\ } (\bibinfo {year} {2021}),\ \Eprint
  {http://arxiv.org/abs/2103.08505} {arXiv:2103.08505 [quant-ph]} \BibitemShut
  {NoStop}%
\bibitem [{\citenamefont {Shen}\ \emph {et~al.}(2017)\citenamefont {Shen},
  \citenamefont {Zhang}, \citenamefont {Zhang}, \citenamefont {Zhang},
  \citenamefont {Yung},\ and\ \citenamefont {Kim}}]{PhysRevA.95.020501}%
  \BibitemOpen
  \bibfield  {author} {\bibinfo {author} {\bibfnamefont {Y.}~\bibnamefont
  {Shen}}, \bibinfo {author} {\bibfnamefont {X.}~\bibnamefont {Zhang}},
  \bibinfo {author} {\bibfnamefont {S.}~\bibnamefont {Zhang}}, \bibinfo
  {author} {\bibfnamefont {J.-N.}\ \bibnamefont {Zhang}}, \bibinfo {author}
  {\bibfnamefont {M.-H.}\ \bibnamefont {Yung}}, \ and\ \bibinfo {author}
  {\bibfnamefont {K.}~\bibnamefont {Kim}},\ }\href {\doibase
  10.1103/PhysRevA.95.020501} {\bibfield  {journal} {\bibinfo  {journal} {Phys.
  Rev. A}\ }\textbf {\bibinfo {volume} {95}},\ \bibinfo {pages} {020501}
  (\bibinfo {year} {2017})}\BibitemShut {NoStop}%
\bibitem [{\citenamefont {McClean}\ \emph {et~al.}(2017)\citenamefont
  {McClean}, \citenamefont {Kimchi-Schwartz}, \citenamefont {Carter},\ and\
  \citenamefont {de~Jong}}]{PhysRevA.95.042308}%
  \BibitemOpen
  \bibfield  {author} {\bibinfo {author} {\bibfnamefont {J.~R.}\ \bibnamefont
  {McClean}}, \bibinfo {author} {\bibfnamefont {M.~E.}\ \bibnamefont
  {Kimchi-Schwartz}}, \bibinfo {author} {\bibfnamefont {J.}~\bibnamefont
  {Carter}}, \ and\ \bibinfo {author} {\bibfnamefont {W.~A.}\ \bibnamefont
  {de~Jong}},\ }\href {\doibase 10.1103/PhysRevA.95.042308} {\bibfield
  {journal} {\bibinfo  {journal} {Phys. Rev. A}\ }\textbf {\bibinfo {volume}
  {95}},\ \bibinfo {pages} {042308} (\bibinfo {year} {2017})}\BibitemShut
  {NoStop}%
\bibitem [{\citenamefont {Nakanishi}\ \emph {et~al.}(2019)\citenamefont
  {Nakanishi}, \citenamefont {Mitarai},\ and\ \citenamefont
  {Fujii}}]{PhysRevResearch.1.033062}%
  \BibitemOpen
  \bibfield  {author} {\bibinfo {author} {\bibfnamefont {K.~M.}\ \bibnamefont
  {Nakanishi}}, \bibinfo {author} {\bibfnamefont {K.}~\bibnamefont {Mitarai}},
  \ and\ \bibinfo {author} {\bibfnamefont {K.}~\bibnamefont {Fujii}},\ }\href
  {\doibase 10.1103/PhysRevResearch.1.033062} {\bibfield  {journal} {\bibinfo
  {journal} {Phys. Rev. Research}\ }\textbf {\bibinfo {volume} {1}},\ \bibinfo
  {pages} {033062} (\bibinfo {year} {2019})}\BibitemShut {NoStop}%
\bibitem [{\citenamefont {Higgott}\ \emph {et~al.}(2019)\citenamefont
  {Higgott}, \citenamefont {Wang},\ and\ \citenamefont
  {Brierley}}]{Higgott_2019}%
  \BibitemOpen
  \bibfield  {author} {\bibinfo {author} {\bibfnamefont {O.}~\bibnamefont
  {Higgott}}, \bibinfo {author} {\bibfnamefont {D.}~\bibnamefont {Wang}}, \
  and\ \bibinfo {author} {\bibfnamefont {S.}~\bibnamefont {Brierley}},\ }\href
  {\doibase 10.22331/q-2019-07-01-156} {\bibfield  {journal} {\bibinfo
  {journal} {Quantum}\ }\textbf {\bibinfo {volume} {3}},\ \bibinfo {pages}
  {156} (\bibinfo {year} {2019})}\BibitemShut {NoStop}%
\bibitem [{\citenamefont {Kawai}\ and\ \citenamefont
  {Nakagawa}(2020)}]{Kawai_2020}%
  \BibitemOpen
  \bibfield  {author} {\bibinfo {author} {\bibfnamefont {H.}~\bibnamefont
  {Kawai}}\ and\ \bibinfo {author} {\bibfnamefont {Y.~O.}\ \bibnamefont
  {Nakagawa}},\ }\href {\doibase 10.1088/2632-2153/aba183} {\bibfield
  {journal} {\bibinfo  {journal} {Machine Learning: Science and Technology}\
  }\textbf {\bibinfo {volume} {1}},\ \bibinfo {pages} {045027} (\bibinfo {year}
  {2020})}\BibitemShut {NoStop}%
\bibitem [{\citenamefont {Greene-Diniz}\ and\ \citenamefont
  {Muñoz~Ramo}(2021)}]{https://doi.org/10.1002/qua.26352}%
  \BibitemOpen
  \bibfield  {author} {\bibinfo {author} {\bibfnamefont {G.}~\bibnamefont
  {Greene-Diniz}}\ and\ \bibinfo {author} {\bibfnamefont {D.}~\bibnamefont
  {Muñoz~Ramo}},\ }\href
  {https://onlinelibrary.wiley.com/doi/abs/10.1002/qua.26352} {\bibfield
  {journal} {\bibinfo  {journal} {Int. J. Quantum Chem.}\ }\textbf {\bibinfo
  {volume} {121}},\ \bibinfo {pages} {e26352} (\bibinfo {year}
  {2021})}\BibitemShut {NoStop}%
\bibitem [{\citenamefont {Kandala}\ \emph {et~al.}(2019)\citenamefont
  {Kandala}, \citenamefont {Temme}, \citenamefont {Córcoles}, \citenamefont
  {Mezzacapo}, \citenamefont {Chow},\ and\ \citenamefont {Gambetta}}]{RN748}%
  \BibitemOpen
  \bibfield  {author} {\bibinfo {author} {\bibfnamefont {A.}~\bibnamefont
  {Kandala}}, \bibinfo {author} {\bibfnamefont {K.}~\bibnamefont {Temme}},
  \bibinfo {author} {\bibfnamefont {A.~D.}\ \bibnamefont {Córcoles}}, \bibinfo
  {author} {\bibfnamefont {A.}~\bibnamefont {Mezzacapo}}, \bibinfo {author}
  {\bibfnamefont {J.~M.}\ \bibnamefont {Chow}}, \ and\ \bibinfo {author}
  {\bibfnamefont {J.~M.}\ \bibnamefont {Gambetta}},\ }\href {\doibase
  10.1038/s41586-019-1040-7} {\bibfield  {journal} {\bibinfo  {journal}
  {Nature}\ }\textbf {\bibinfo {volume} {567}},\ \bibinfo {pages} {491}
  (\bibinfo {year} {2019})}\BibitemShut {NoStop}%
\bibitem [{\citenamefont {Kandala}\ \emph {et~al.}(2017)\citenamefont
  {Kandala}, \citenamefont {Mezzacapo}, \citenamefont {Temme}, \citenamefont
  {Takita}, \citenamefont {Brink}, \citenamefont {Chow},\ and\ \citenamefont
  {Gambetta}}]{Kandala2017Nature}%
  \BibitemOpen
  \bibfield  {author} {\bibinfo {author} {\bibfnamefont {A.}~\bibnamefont
  {Kandala}}, \bibinfo {author} {\bibfnamefont {A.}~\bibnamefont {Mezzacapo}},
  \bibinfo {author} {\bibfnamefont {K.}~\bibnamefont {Temme}}, \bibinfo
  {author} {\bibfnamefont {M.}~\bibnamefont {Takita}}, \bibinfo {author}
  {\bibfnamefont {M.}~\bibnamefont {Brink}}, \bibinfo {author} {\bibfnamefont
  {J.~M.}\ \bibnamefont {Chow}}, \ and\ \bibinfo {author} {\bibfnamefont
  {J.~M.}\ \bibnamefont {Gambetta}},\ }\href
  {https://www.ncbi.nlm.nih.gov/pubmed/28905916} {\bibfield  {journal}
  {\bibinfo  {journal} {Nature}\ }\textbf {\bibinfo {volume} {549}},\ \bibinfo
  {pages} {242} (\bibinfo {year} {2017})}\BibitemShut {NoStop}%
\bibitem [{\citenamefont {Tang}\ \emph
  {et~al.}(2021{\natexlab{a}})\citenamefont {Tang}, \citenamefont {Shkolnikov},
  \citenamefont {Barron}, \citenamefont {Grimsley}, \citenamefont {Mayhall},
  \citenamefont {Barnes},\ and\ \citenamefont {Economou}}]{RN671}%
  \BibitemOpen
  \bibfield  {author} {\bibinfo {author} {\bibfnamefont {H.~L.}\ \bibnamefont
  {Tang}}, \bibinfo {author} {\bibfnamefont {V.}~\bibnamefont {Shkolnikov}},
  \bibinfo {author} {\bibfnamefont {G.~S.}\ \bibnamefont {Barron}}, \bibinfo
  {author} {\bibfnamefont {H.~R.}\ \bibnamefont {Grimsley}}, \bibinfo {author}
  {\bibfnamefont {N.~J.}\ \bibnamefont {Mayhall}}, \bibinfo {author}
  {\bibfnamefont {E.}~\bibnamefont {Barnes}}, \ and\ \bibinfo {author}
  {\bibfnamefont {S.~E.}\ \bibnamefont {Economou}},\ }\href {\doibase
  10.1103/PRXQuantum.2.020310} {\bibfield  {journal} {\bibinfo  {journal} {PRX
  Quantum}\ }\textbf {\bibinfo {volume} {2}},\ \bibinfo {pages} {020310}
  (\bibinfo {year} {2021}{\natexlab{a}})}\BibitemShut {NoStop}%
\bibitem [{\citenamefont {McClean}\ \emph {et~al.}(2018)\citenamefont
  {McClean}, \citenamefont {Boixo}, \citenamefont {Smelyanskiy}, \citenamefont
  {Babbush},\ and\ \citenamefont {Neven}}]{RN861}%
  \BibitemOpen
  \bibfield  {author} {\bibinfo {author} {\bibfnamefont {J.~R.}\ \bibnamefont
  {McClean}}, \bibinfo {author} {\bibfnamefont {S.}~\bibnamefont {Boixo}},
  \bibinfo {author} {\bibfnamefont {V.~N.}\ \bibnamefont {Smelyanskiy}},
  \bibinfo {author} {\bibfnamefont {R.}~\bibnamefont {Babbush}}, \ and\
  \bibinfo {author} {\bibfnamefont {H.}~\bibnamefont {Neven}},\ }\href
  {\doibase 10.1038/s41467-018-07090-4} {\bibfield  {journal} {\bibinfo
  {journal} {Nat. Commun.}\ }\textbf {\bibinfo {volume} {9}},\ \bibinfo {pages}
  {4812} (\bibinfo {year} {2018})}\BibitemShut {NoStop}%
\bibitem [{\citenamefont {Wang}\ \emph {et~al.}(2021)\citenamefont {Wang},
  \citenamefont {Fontana}, \citenamefont {Cerezo}, \citenamefont {Sharma},
  \citenamefont {Sone}, \citenamefont {Cincio},\ and\ \citenamefont
  {Coles}}]{wang2021noiseinduced}%
  \BibitemOpen
  \bibfield  {author} {\bibinfo {author} {\bibfnamefont {S.}~\bibnamefont
  {Wang}}, \bibinfo {author} {\bibfnamefont {E.}~\bibnamefont {Fontana}},
  \bibinfo {author} {\bibfnamefont {M.}~\bibnamefont {Cerezo}}, \bibinfo
  {author} {\bibfnamefont {K.}~\bibnamefont {Sharma}}, \bibinfo {author}
  {\bibfnamefont {A.}~\bibnamefont {Sone}}, \bibinfo {author} {\bibfnamefont
  {L.}~\bibnamefont {Cincio}}, \ and\ \bibinfo {author} {\bibfnamefont {P.~J.}\
  \bibnamefont {Coles}},\ }\href@noop {} {\enquote {\bibinfo {title}
  {Noise-induced barren plateaus in variational quantum algorithms},}\ }
  (\bibinfo {year} {2021}),\ \Eprint {http://arxiv.org/abs/2007.14384}
  {arXiv:2007.14384 [quant-ph]} \BibitemShut {NoStop}%
\bibitem [{\citenamefont {Cerezo}\ \emph {et~al.}(2021)\citenamefont {Cerezo},
  \citenamefont {Sone}, \citenamefont {Volkoff}, \citenamefont {Cincio},\ and\
  \citenamefont {Coles}}]{RN670}%
  \BibitemOpen
  \bibfield  {author} {\bibinfo {author} {\bibfnamefont {M.}~\bibnamefont
  {Cerezo}}, \bibinfo {author} {\bibfnamefont {A.}~\bibnamefont {Sone}},
  \bibinfo {author} {\bibfnamefont {T.}~\bibnamefont {Volkoff}}, \bibinfo
  {author} {\bibfnamefont {L.}~\bibnamefont {Cincio}}, \ and\ \bibinfo {author}
  {\bibfnamefont {P.~J.}\ \bibnamefont {Coles}},\ }\href {\doibase
  10.1038/s41467-021-21728-w} {\bibfield  {journal} {\bibinfo  {journal} {Nat.
  Commun.}\ }\textbf {\bibinfo {volume} {12}},\ \bibinfo {pages} {1791}
  (\bibinfo {year} {2021})}\BibitemShut {NoStop}%
\bibitem [{\citenamefont {Abbas}\ \emph {et~al.}(2020)\citenamefont {Abbas},
  \citenamefont {Sutter}, \citenamefont {Zoufal}, \citenamefont {Lucchi},
  \citenamefont {Figalli},\ and\ \citenamefont {Woerner}}]{abbas2020power}%
  \BibitemOpen
  \bibfield  {author} {\bibinfo {author} {\bibfnamefont {A.}~\bibnamefont
  {Abbas}}, \bibinfo {author} {\bibfnamefont {D.}~\bibnamefont {Sutter}},
  \bibinfo {author} {\bibfnamefont {C.}~\bibnamefont {Zoufal}}, \bibinfo
  {author} {\bibfnamefont {A.}~\bibnamefont {Lucchi}}, \bibinfo {author}
  {\bibfnamefont {A.}~\bibnamefont {Figalli}}, \ and\ \bibinfo {author}
  {\bibfnamefont {S.}~\bibnamefont {Woerner}},\ }\href@noop {} {\enquote
  {\bibinfo {title} {The power of quantum neural networks},}\ } (\bibinfo
  {year} {2020}),\ \Eprint {http://arxiv.org/abs/2011.00027} {arXiv:2011.00027
  [quant-ph]} \BibitemShut {NoStop}%
\bibitem [{\citenamefont {Bartlett}\ \emph {et~al.}(1989)\citenamefont
  {Bartlett}, \citenamefont {Kucharski},\ and\ \citenamefont
  {Noga}}]{BARTLETT1989133}%
  \BibitemOpen
  \bibfield  {author} {\bibinfo {author} {\bibfnamefont {R.~J.}\ \bibnamefont
  {Bartlett}}, \bibinfo {author} {\bibfnamefont {S.~A.}\ \bibnamefont
  {Kucharski}}, \ and\ \bibinfo {author} {\bibfnamefont {J.}~\bibnamefont
  {Noga}},\ }\href {\doibase https://doi.org/10.1016/S0009-2614(89)87372-5}
  {\bibfield  {journal} {\bibinfo  {journal} {Chem. Phys. Lett.}\ }\textbf
  {\bibinfo {volume} {155}},\ \bibinfo {pages} {133 } (\bibinfo {year}
  {1989})}\BibitemShut {NoStop}%
\bibitem [{\citenamefont {Bartlett}\ and\ \citenamefont
  {Musia\l{}}(2007)}]{RevModPhys.79.291}%
  \BibitemOpen
  \bibfield  {author} {\bibinfo {author} {\bibfnamefont {R.~J.}\ \bibnamefont
  {Bartlett}}\ and\ \bibinfo {author} {\bibfnamefont {M.}~\bibnamefont
  {Musia\l{}}},\ }\href {\doibase 10.1103/RevModPhys.79.291} {\bibfield
  {journal} {\bibinfo  {journal} {Rev. Mod. Phys.}\ }\textbf {\bibinfo {volume}
  {79}},\ \bibinfo {pages} {291} (\bibinfo {year} {2007})}\BibitemShut
  {NoStop}%
\bibitem [{\citenamefont {Seeley}\ \emph
  {et~al.}(2012{\natexlab{a}})\citenamefont {Seeley}, \citenamefont {Richard},\
  and\ \citenamefont {Love}}]{jcp4768229}%
  \BibitemOpen
  \bibfield  {author} {\bibinfo {author} {\bibfnamefont {J.~T.}\ \bibnamefont
  {Seeley}}, \bibinfo {author} {\bibfnamefont {M.~J.}\ \bibnamefont {Richard}},
  \ and\ \bibinfo {author} {\bibfnamefont {P.~J.}\ \bibnamefont {Love}},\
  }\href {\doibase 10.1063/1.4768229} {\bibfield  {journal} {\bibinfo
  {journal} {J. Chem. Phys.}\ }\textbf {\bibinfo {volume} {137}},\ \bibinfo
  {pages} {224109} (\bibinfo {year} {2012}{\natexlab{a}})}\BibitemShut
  {NoStop}%
\bibitem [{\citenamefont {Barkoutsos}\ \emph {et~al.}(2018)\citenamefont
  {Barkoutsos}, \citenamefont {Gonthier}, \citenamefont {Sokolov},
  \citenamefont {Moll}, \citenamefont {Salis}, \citenamefont {Fuhrer},
  \citenamefont {Ganzhorn}, \citenamefont {Egger}, \citenamefont {Troyer},
  \citenamefont {Mezzacapo}, \citenamefont {Filipp},\ and\ \citenamefont
  {Tavernelli}}]{PhysRevA.98.022322}%
  \BibitemOpen
  \bibfield  {author} {\bibinfo {author} {\bibfnamefont {P.~K.}\ \bibnamefont
  {Barkoutsos}}, \bibinfo {author} {\bibfnamefont {J.~F.}\ \bibnamefont
  {Gonthier}}, \bibinfo {author} {\bibfnamefont {I.}~\bibnamefont {Sokolov}},
  \bibinfo {author} {\bibfnamefont {N.}~\bibnamefont {Moll}}, \bibinfo {author}
  {\bibfnamefont {G.}~\bibnamefont {Salis}}, \bibinfo {author} {\bibfnamefont
  {A.}~\bibnamefont {Fuhrer}}, \bibinfo {author} {\bibfnamefont
  {M.}~\bibnamefont {Ganzhorn}}, \bibinfo {author} {\bibfnamefont {D.~J.}\
  \bibnamefont {Egger}}, \bibinfo {author} {\bibfnamefont {M.}~\bibnamefont
  {Troyer}}, \bibinfo {author} {\bibfnamefont {A.}~\bibnamefont {Mezzacapo}},
  \bibinfo {author} {\bibfnamefont {S.}~\bibnamefont {Filipp}}, \ and\ \bibinfo
  {author} {\bibfnamefont {I.}~\bibnamefont {Tavernelli}},\ }\href {\doibase
  10.1103/PhysRevA.98.022322} {\bibfield  {journal} {\bibinfo  {journal} {Phys.
  Rev. A}\ }\textbf {\bibinfo {volume} {98}},\ \bibinfo {pages} {022322}
  (\bibinfo {year} {2018})}\BibitemShut {NoStop}%
\bibitem [{\citenamefont {Romero}\ \emph {et~al.}(2018)\citenamefont {Romero},
  \citenamefont {Babbush}, \citenamefont {McClean}, \citenamefont {Hempel},
  \citenamefont {Love},\ and\ \citenamefont {Aspuru-Guzik}}]{Alan2018QST}%
  \BibitemOpen
  \bibfield  {author} {\bibinfo {author} {\bibfnamefont {J.}~\bibnamefont
  {Romero}}, \bibinfo {author} {\bibfnamefont {R.}~\bibnamefont {Babbush}},
  \bibinfo {author} {\bibfnamefont {J.}~\bibnamefont {McClean}}, \bibinfo
  {author} {\bibfnamefont {C.}~\bibnamefont {Hempel}}, \bibinfo {author}
  {\bibfnamefont {P.~J.}\ \bibnamefont {Love}}, \ and\ \bibinfo {author}
  {\bibfnamefont {A.}~\bibnamefont {Aspuru-Guzik}},\ }\href
  {http://iopscience.iop.org/article/10.1088/2058-9565/aad3e4} {\bibfield
  {journal} {\bibinfo  {journal} {Quantum Sci. Technol.}\ }\textbf {\bibinfo
  {volume} {4}},\ \bibinfo {pages} {14008} (\bibinfo {year}
  {2018})}\BibitemShut {NoStop}%
\bibitem [{\citenamefont {Kühn}\ \emph {et~al.}(2019)\citenamefont {Kühn},
  \citenamefont {Zanker}, \citenamefont {Deglmann}, \citenamefont {Marthaler},\
  and\ \citenamefont {Weiß}}]{acs.jctc.9b00236}%
  \BibitemOpen
  \bibfield  {author} {\bibinfo {author} {\bibfnamefont {M.}~\bibnamefont
  {Kühn}}, \bibinfo {author} {\bibfnamefont {S.}~\bibnamefont {Zanker}},
  \bibinfo {author} {\bibfnamefont {P.}~\bibnamefont {Deglmann}}, \bibinfo
  {author} {\bibfnamefont {M.}~\bibnamefont {Marthaler}}, \ and\ \bibinfo
  {author} {\bibfnamefont {H.}~\bibnamefont {Weiß}},\ }\href {\doibase
  10.1021/acs.jctc.9b00236} {\bibfield  {journal} {\bibinfo  {journal} {J.
  Chem. Theory Comput.}\ }\textbf {\bibinfo {volume} {15}},\ \bibinfo {pages}
  {4764} (\bibinfo {year} {2019})},\ \bibinfo {note} {pMID:
  31403781}\BibitemShut {NoStop}%
\bibitem [{\citenamefont {Ryabinkin}\ \emph {et~al.}(2018)\citenamefont
  {Ryabinkin}, \citenamefont {Yen}, \citenamefont {Genin},\ and\ \citenamefont
  {Izmaylov}}]{jctc8b00932}%
  \BibitemOpen
  \bibfield  {author} {\bibinfo {author} {\bibfnamefont {I.~G.}\ \bibnamefont
  {Ryabinkin}}, \bibinfo {author} {\bibfnamefont {T.-C.}\ \bibnamefont {Yen}},
  \bibinfo {author} {\bibfnamefont {S.~N.}\ \bibnamefont {Genin}}, \ and\
  \bibinfo {author} {\bibfnamefont {A.~F.}\ \bibnamefont {Izmaylov}},\ }\href
  {\doibase 10.1021/acs.jctc.8b00932} {\bibfield  {journal} {\bibinfo
  {journal} {J. Chem. Theory Comput.}\ }\textbf {\bibinfo {volume} {14}},\
  \bibinfo {pages} {6317} (\bibinfo {year} {2018})}\BibitemShut {NoStop}%
\bibitem [{\citenamefont {Lee}\ \emph {et~al.}(2019)\citenamefont {Lee},
  \citenamefont {Huggins}, \citenamefont {Head-Gordon},\ and\ \citenamefont
  {Whaley}}]{jctc.8b01004}%
  \BibitemOpen
  \bibfield  {author} {\bibinfo {author} {\bibfnamefont {J.}~\bibnamefont
  {Lee}}, \bibinfo {author} {\bibfnamefont {W.~J.}\ \bibnamefont {Huggins}},
  \bibinfo {author} {\bibfnamefont {M.}~\bibnamefont {Head-Gordon}}, \ and\
  \bibinfo {author} {\bibfnamefont {K.~B.}\ \bibnamefont {Whaley}},\ }\href
  {\doibase 10.1021/acs.jctc.8b01004} {\bibfield  {journal} {\bibinfo
  {journal} {J. Chem. Theory Comput.}\ }\textbf {\bibinfo {volume} {15}},\
  \bibinfo {pages} {311} (\bibinfo {year} {2019})}\BibitemShut {NoStop}%
\bibitem [{\citenamefont {Matsuzawa}\ and\ \citenamefont
  {Kurashige}(2020)}]{jctc9b00963}%
  \BibitemOpen
  \bibfield  {author} {\bibinfo {author} {\bibfnamefont {Y.}~\bibnamefont
  {Matsuzawa}}\ and\ \bibinfo {author} {\bibfnamefont {Y.}~\bibnamefont
  {Kurashige}},\ }\href {\doibase 10.1021/acs.jctc.9b00963} {\bibfield
  {journal} {\bibinfo  {journal} {J. Chem. Theory Comput.}\ }\textbf {\bibinfo
  {volume} {16}},\ \bibinfo {pages} {944} (\bibinfo {year} {2020})}\BibitemShut
  {NoStop}%
\bibitem [{\citenamefont {Xia}\ and\ \citenamefont {Kais}(2020)}]{Xia_2020}%
  \BibitemOpen
  \bibfield  {author} {\bibinfo {author} {\bibfnamefont {R.}~\bibnamefont
  {Xia}}\ and\ \bibinfo {author} {\bibfnamefont {S.}~\bibnamefont {Kais}},\
  }\href {\doibase 10.1088/2058-9565/abbc74} {\bibfield  {journal} {\bibinfo
  {journal} {Quantum Sci. Tech.}\ }\textbf {\bibinfo {volume} {6}},\ \bibinfo
  {pages} {015001} (\bibinfo {year} {2020})}\BibitemShut {NoStop}%
\bibitem [{\citenamefont {Tkachenko}\ \emph {et~al.}(2021)\citenamefont
  {Tkachenko}, \citenamefont {Sud}, \citenamefont {Zhang}, \citenamefont
  {Tretiak}, \citenamefont {Anisimov}, \citenamefont {Arrasmith}, \citenamefont
  {Coles}, \citenamefont {Cincio},\ and\ \citenamefont {Dub}}]{permVQE2021}%
  \BibitemOpen
  \bibfield  {author} {\bibinfo {author} {\bibfnamefont {N.~V.}\ \bibnamefont
  {Tkachenko}}, \bibinfo {author} {\bibfnamefont {J.}~\bibnamefont {Sud}},
  \bibinfo {author} {\bibfnamefont {Y.}~\bibnamefont {Zhang}}, \bibinfo
  {author} {\bibfnamefont {S.}~\bibnamefont {Tretiak}}, \bibinfo {author}
  {\bibfnamefont {P.~M.}\ \bibnamefont {Anisimov}}, \bibinfo {author}
  {\bibfnamefont {A.~T.}\ \bibnamefont {Arrasmith}}, \bibinfo {author}
  {\bibfnamefont {P.~J.}\ \bibnamefont {Coles}}, \bibinfo {author}
  {\bibfnamefont {L.}~\bibnamefont {Cincio}}, \ and\ \bibinfo {author}
  {\bibfnamefont {P.~A.}\ \bibnamefont {Dub}},\ }\href
  {https://journals.aps.org/prxquantum/abstract/10.1103/PRXQuantum.2.020337}
  {\bibfield  {journal} {\bibinfo  {journal} {PRX Quantum}\ }\textbf {\bibinfo
  {volume} {2}},\ \bibinfo {pages} {020337} (\bibinfo {year}
  {2021})}\BibitemShut {NoStop}%
\bibitem [{\citenamefont {Grimsley}\ \emph {et~al.}(2019)\citenamefont
  {Grimsley}, \citenamefont {Economou}, \citenamefont {Barnes},\ and\
  \citenamefont {Mayhall}}]{RN170}%
  \BibitemOpen
  \bibfield  {author} {\bibinfo {author} {\bibfnamefont {H.~R.}\ \bibnamefont
  {Grimsley}}, \bibinfo {author} {\bibfnamefont {S.~E.}\ \bibnamefont
  {Economou}}, \bibinfo {author} {\bibfnamefont {E.}~\bibnamefont {Barnes}}, \
  and\ \bibinfo {author} {\bibfnamefont {N.~J.}\ \bibnamefont {Mayhall}},\
  }\href {\doibase 10.1038/s41467-019-10988-2} {\bibfield  {journal} {\bibinfo
  {journal} {Nat. Commun.}\ }\textbf {\bibinfo {volume} {10}},\ \bibinfo
  {pages} {3007} (\bibinfo {year} {2019})}\BibitemShut {NoStop}%
\bibitem [{\citenamefont {Tang}\ \emph
  {et~al.}(2021{\natexlab{b}})\citenamefont {Tang}, \citenamefont {Shkolnikov},
  \citenamefont {Barron}, \citenamefont {Grimsley}, \citenamefont {Mayhall},
  \citenamefont {Barnes},\ and\ \citenamefont
  {Economou}}]{PRXQuantum.2.020310}%
  \BibitemOpen
  \bibfield  {author} {\bibinfo {author} {\bibfnamefont {H.~L.}\ \bibnamefont
  {Tang}}, \bibinfo {author} {\bibfnamefont {V.}~\bibnamefont {Shkolnikov}},
  \bibinfo {author} {\bibfnamefont {G.~S.}\ \bibnamefont {Barron}}, \bibinfo
  {author} {\bibfnamefont {H.~R.}\ \bibnamefont {Grimsley}}, \bibinfo {author}
  {\bibfnamefont {N.~J.}\ \bibnamefont {Mayhall}}, \bibinfo {author}
  {\bibfnamefont {E.}~\bibnamefont {Barnes}}, \ and\ \bibinfo {author}
  {\bibfnamefont {S.~E.}\ \bibnamefont {Economou}},\ }\href {\doibase
  10.1103/PRXQuantum.2.020310} {\bibfield  {journal} {\bibinfo  {journal} {PRX
  Quantum}\ }\textbf {\bibinfo {volume} {2}},\ \bibinfo {pages} {020310}
  (\bibinfo {year} {2021}{\natexlab{b}})}\BibitemShut {NoStop}%
\bibitem [{\citenamefont {Ryabinkin}\ \emph {et~al.}(2020)\citenamefont
  {Ryabinkin}, \citenamefont {Lang}, \citenamefont {Genin},\ and\ \citenamefont
  {Izmaylov}}]{jctc9b01084}%
  \BibitemOpen
  \bibfield  {author} {\bibinfo {author} {\bibfnamefont {I.~G.}\ \bibnamefont
  {Ryabinkin}}, \bibinfo {author} {\bibfnamefont {R.~A.}\ \bibnamefont {Lang}},
  \bibinfo {author} {\bibfnamefont {S.~N.}\ \bibnamefont {Genin}}, \ and\
  \bibinfo {author} {\bibfnamefont {A.~F.}\ \bibnamefont {Izmaylov}},\ }\href
  {https://doi.org/10.1021/acs.jctc.9b01084} {\bibfield  {journal} {\bibinfo
  {journal} {J. Chem. Theory Comput.}\ }\textbf {\bibinfo {volume} {16}},\
  \bibinfo {pages} {1055} (\bibinfo {year} {2020})}\BibitemShut {NoStop}%
\bibitem [{\citenamefont {Lang}\ \emph {et~al.}(2021)\citenamefont {Lang},
  \citenamefont {Ryabinkin},\ and\ \citenamefont {Izmaylov}}]{RN669}%
  \BibitemOpen
  \bibfield  {author} {\bibinfo {author} {\bibfnamefont {R.~A.}\ \bibnamefont
  {Lang}}, \bibinfo {author} {\bibfnamefont {I.~G.}\ \bibnamefont {Ryabinkin}},
  \ and\ \bibinfo {author} {\bibfnamefont {A.~F.}\ \bibnamefont {Izmaylov}},\
  }\href {\doibase 10.1021/acs.jctc.0c00170} {\bibfield  {journal} {\bibinfo
  {journal} {J. Chem. Theory Comput.}\ }\textbf {\bibinfo {volume} {17}},\
  \bibinfo {pages} {66} (\bibinfo {year} {2021})}\BibitemShut {NoStop}%
\bibitem [{\citenamefont {Ryabinkin}\ \emph {et~al.}(2021)\citenamefont
  {Ryabinkin}, \citenamefont {Izmaylov},\ and\ \citenamefont
  {Genin}}]{Ryabinkin_2021}%
  \BibitemOpen
  \bibfield  {author} {\bibinfo {author} {\bibfnamefont {I.~G.}\ \bibnamefont
  {Ryabinkin}}, \bibinfo {author} {\bibfnamefont {A.~F.}\ \bibnamefont
  {Izmaylov}}, \ and\ \bibinfo {author} {\bibfnamefont {S.~N.}\ \bibnamefont
  {Genin}},\ }\href {\doibase 10.1088/2058-9565/abda8e} {\bibfield  {journal}
  {\bibinfo  {journal} {Quantum Sci. Tech.}\ }\textbf {\bibinfo {volume} {6}},\
  \bibinfo {pages} {024012} (\bibinfo {year} {2021})}\BibitemShut {NoStop}%
\bibitem [{\citenamefont {Yordanov}\ \emph {et~al.}(2020)\citenamefont
  {Yordanov}, \citenamefont {Armaos}, \citenamefont {Barnes},\ and\
  \citenamefont {Arvidsson-Shukur}}]{yordanov2020iterative}%
  \BibitemOpen
  \bibfield  {author} {\bibinfo {author} {\bibfnamefont {Y.~S.}\ \bibnamefont
  {Yordanov}}, \bibinfo {author} {\bibfnamefont {V.}~\bibnamefont {Armaos}},
  \bibinfo {author} {\bibfnamefont {C.~H.~W.}\ \bibnamefont {Barnes}}, \ and\
  \bibinfo {author} {\bibfnamefont {D.~R.~M.}\ \bibnamefont
  {Arvidsson-Shukur}},\ }\href@noop {} {\enquote {\bibinfo {title} {Iterative
  qubit-excitation based variational quantum eigensolver},}\ } (\bibinfo {year}
  {2020}),\ \Eprint {http://arxiv.org/abs/2011.10540} {arXiv:2011.10540
  [quant-ph]} \BibitemShut {NoStop}%
\bibitem [{\citenamefont {Bravyi}\ \emph {et~al.}(2017)\citenamefont {Bravyi},
  \citenamefont {Gambetta}, \citenamefont {Mezzacapo},\ and\ \citenamefont
  {Temme}}]{bravyi2017tapering}%
  \BibitemOpen
  \bibfield  {author} {\bibinfo {author} {\bibfnamefont {S.}~\bibnamefont
  {Bravyi}}, \bibinfo {author} {\bibfnamefont {J.~M.}\ \bibnamefont
  {Gambetta}}, \bibinfo {author} {\bibfnamefont {A.}~\bibnamefont {Mezzacapo}},
  \ and\ \bibinfo {author} {\bibfnamefont {K.}~\bibnamefont {Temme}},\
  }\href@noop {} {\enquote {\bibinfo {title} {Tapering off qubits to simulate
  fermionic hamiltonians},}\ } (\bibinfo {year} {2017}),\ \Eprint
  {http://arxiv.org/abs/1701.08213} {arXiv:1701.08213 [quant-ph]} \BibitemShut
  {NoStop}%
\bibitem [{\citenamefont {Setia}\ \emph {et~al.}(2020)\citenamefont {Setia},
  \citenamefont {Chen}, \citenamefont {Rice}, \citenamefont {Mezzacapo},
  \citenamefont {Pistoia},\ and\ \citenamefont {Whitfield}}]{RN676}%
  \BibitemOpen
  \bibfield  {author} {\bibinfo {author} {\bibfnamefont {K.}~\bibnamefont
  {Setia}}, \bibinfo {author} {\bibfnamefont {R.}~\bibnamefont {Chen}},
  \bibinfo {author} {\bibfnamefont {J.~E.}\ \bibnamefont {Rice}}, \bibinfo
  {author} {\bibfnamefont {A.}~\bibnamefont {Mezzacapo}}, \bibinfo {author}
  {\bibfnamefont {M.}~\bibnamefont {Pistoia}}, \ and\ \bibinfo {author}
  {\bibfnamefont {J.~D.}\ \bibnamefont {Whitfield}},\ }\href {\doibase
  10.1021/acs.jctc.0c00113} {\bibfield  {journal} {\bibinfo  {journal} {J.
  Chem. Theory Comput.}\ }\textbf {\bibinfo {volume} {16}},\ \bibinfo {pages}
  {6091} (\bibinfo {year} {2020})}\BibitemShut {NoStop}%
\bibitem [{\citenamefont {Yen}\ \emph {et~al.}(2019)\citenamefont {Yen},
  \citenamefont {Lang},\ and\ \citenamefont
  {Izmaylov}}]{doi:10.1063/1.5110682}%
  \BibitemOpen
  \bibfield  {author} {\bibinfo {author} {\bibfnamefont {T.-C.}\ \bibnamefont
  {Yen}}, \bibinfo {author} {\bibfnamefont {R.~A.}\ \bibnamefont {Lang}}, \
  and\ \bibinfo {author} {\bibfnamefont {A.~F.}\ \bibnamefont {Izmaylov}},\
  }\href {\doibase 10.1063/1.5110682} {\bibfield  {journal} {\bibinfo
  {journal} {J. Chem. Phys.}\ }\textbf {\bibinfo {volume} {151}},\ \bibinfo
  {pages} {164111} (\bibinfo {year} {2019})},\ \Eprint
  {http://arxiv.org/abs/https://doi.org/10.1063/1.5110682}
  {https://doi.org/10.1063/1.5110682} \BibitemShut {NoStop}%
\bibitem [{\citenamefont {Zhang}\ \emph
  {et~al.}(2021{\natexlab{a}})\citenamefont {Zhang}, \citenamefont {Gomes},
  \citenamefont {Berthusen}, \citenamefont {Orth}, \citenamefont {Wang},
  \citenamefont {Ho},\ and\ \citenamefont {Yao}}]{PhysRevResearch.3.013039}%
  \BibitemOpen
  \bibfield  {author} {\bibinfo {author} {\bibfnamefont {F.}~\bibnamefont
  {Zhang}}, \bibinfo {author} {\bibfnamefont {N.}~\bibnamefont {Gomes}},
  \bibinfo {author} {\bibfnamefont {N.~F.}\ \bibnamefont {Berthusen}}, \bibinfo
  {author} {\bibfnamefont {P.~P.}\ \bibnamefont {Orth}}, \bibinfo {author}
  {\bibfnamefont {C.-Z.}\ \bibnamefont {Wang}}, \bibinfo {author}
  {\bibfnamefont {K.-M.}\ \bibnamefont {Ho}}, \ and\ \bibinfo {author}
  {\bibfnamefont {Y.-X.}\ \bibnamefont {Yao}},\ }\href {\doibase
  10.1103/PhysRevResearch.3.013039} {\bibfield  {journal} {\bibinfo  {journal}
  {Phys. Rev. Research}\ }\textbf {\bibinfo {volume} {3}},\ \bibinfo {pages}
  {013039} (\bibinfo {year} {2021}{\natexlab{a}})}\BibitemShut {NoStop}%
\bibitem [{\citenamefont {Rissler}\ \emph {et~al.}(2006)\citenamefont
  {Rissler}, \citenamefont {Noack},\ and\ \citenamefont {White}}]{Rissler2006}%
  \BibitemOpen
  \bibfield  {author} {\bibinfo {author} {\bibfnamefont {J.}~\bibnamefont
  {Rissler}}, \bibinfo {author} {\bibfnamefont {R.~M.}\ \bibnamefont {Noack}},
  \ and\ \bibinfo {author} {\bibfnamefont {S.~R.}\ \bibnamefont {White}},\
  }\href {https://www.sciencedirect.com/science/article/pii/S0301010405005069}
  {\bibfield  {journal} {\bibinfo  {journal} {Chem. Phys}\ }\textbf {\bibinfo
  {volume} {323}},\ \bibinfo {pages} {519} (\bibinfo {year}
  {2006})}\BibitemShut {NoStop}%
\bibitem [{\citenamefont {Huang}\ and\ \citenamefont
  {Kais}(2005)}]{HUANG20051}%
  \BibitemOpen
  \bibfield  {author} {\bibinfo {author} {\bibfnamefont {Z.}~\bibnamefont
  {Huang}}\ and\ \bibinfo {author} {\bibfnamefont {S.}~\bibnamefont {Kais}},\
  }\href {\doibase https://doi.org/10.1016/j.cplett.2005.07.045} {\bibfield
  {journal} {\bibinfo  {journal} {Chem. Phys. Lett.}\ }\textbf {\bibinfo
  {volume} {413}},\ \bibinfo {pages} {1 } (\bibinfo {year} {2005})}\BibitemShut
  {NoStop}%
\bibitem [{\citenamefont {Zhang}\ \emph
  {et~al.}(2021{\natexlab{b}})\citenamefont {Zhang}, \citenamefont {Kyaw},
  \citenamefont {Kottmann}, \citenamefont {Degroote},\ and\ \citenamefont
  {Aspuru-Guzik}}]{RN854}%
  \BibitemOpen
  \bibfield  {author} {\bibinfo {author} {\bibfnamefont {Z.-J.}\ \bibnamefont
  {Zhang}}, \bibinfo {author} {\bibfnamefont {T.~H.}\ \bibnamefont {Kyaw}},
  \bibinfo {author} {\bibfnamefont {J.}~\bibnamefont {Kottmann}}, \bibinfo
  {author} {\bibfnamefont {M.}~\bibnamefont {Degroote}}, \ and\ \bibinfo
  {author} {\bibfnamefont {A.}~\bibnamefont {Aspuru-Guzik}},\ }\href
  {http://iopscience.iop.org/article/10.1088/2058-9565/abdca4} {\bibfield
  {journal} {\bibinfo  {journal} {Quantum Sci.Tech.}\ } (\bibinfo {year}
  {2021}{\natexlab{b}})}\BibitemShut {NoStop}%
\bibitem [{\citenamefont {https://metis.readthedocs.io/en/latest}()}]{metis}%
  \BibitemOpen
  \bibfield  {author} {\bibinfo {author} {\bibnamefont
  {https://metis.readthedocs.io/en/latest}},\ }\href
  {https://metis.readthedocs.io/en/latest/} {\enquote {\bibinfo {title} {Metis
  for python},}\ }\BibitemShut {NoStop}%
\bibitem [{\citenamefont {Mniszewski}\ \emph {et~al.}(2021)\citenamefont
  {Mniszewski}, \citenamefont {Dub}, \citenamefont {Tretiak}, \citenamefont
  {Anisimov}, \citenamefont {Zhang},\ and\ \citenamefont
  {Negre}}]{sue_qcd2021}%
  \BibitemOpen
  \bibfield  {author} {\bibinfo {author} {\bibfnamefont {S.~M.}\ \bibnamefont
  {Mniszewski}}, \bibinfo {author} {\bibfnamefont {P.~A.}\ \bibnamefont {Dub}},
  \bibinfo {author} {\bibfnamefont {S.}~\bibnamefont {Tretiak}}, \bibinfo
  {author} {\bibfnamefont {P.~M.}\ \bibnamefont {Anisimov}}, \bibinfo {author}
  {\bibfnamefont {Y.}~\bibnamefont {Zhang}}, \ and\ \bibinfo {author}
  {\bibfnamefont {C.~F.~A.}\ \bibnamefont {Negre}},\ }\href {\doibase
  10.1038/s41598-021-83561-x} {\bibfield  {journal} {\bibinfo  {journal} {Sci.
  Rep.}\ }\textbf {\bibinfo {volume} {11}},\ \bibinfo {pages} {4099} (\bibinfo
  {year} {2021})}\BibitemShut {NoStop}%
\bibitem [{\citenamefont {Booth}\ \emph {et~al.}(2017)\citenamefont {Booth},
  \citenamefont {Reinhardt},\ and\ \citenamefont {Roy}}]{qbsolve}%
  \BibitemOpen
  \bibfield  {author} {\bibinfo {author} {\bibfnamefont {M.}~\bibnamefont
  {Booth}}, \bibinfo {author} {\bibfnamefont {S.~P.}\ \bibnamefont
  {Reinhardt}}, \ and\ \bibinfo {author} {\bibfnamefont {A.}~\bibnamefont
  {Roy}},\ }\href@noop {} {\emph {\bibinfo {title} {Partitioning Optimization
  Problems for Hybrid Classical / Quantum Execution}}},\ \bibinfo {type} {Tech.
  Rep.}\ (\bibinfo {year} {2017})\BibitemShut {NoStop}%
\bibitem [{\citenamefont {Jordan}\ and\ \citenamefont {Wigner}(1928)}]{RN126}%
  \BibitemOpen
  \bibfield  {author} {\bibinfo {author} {\bibfnamefont {P.}~\bibnamefont
  {Jordan}}\ and\ \bibinfo {author} {\bibfnamefont {E.}~\bibnamefont
  {Wigner}},\ }\href {\doibase 10.1007/BF01331938} {\bibfield  {journal}
  {\bibinfo  {journal} {Zeitschrift für Physik}\ }\textbf {\bibinfo {volume}
  {47}},\ \bibinfo {pages} {631} (\bibinfo {year} {1928})}\BibitemShut
  {NoStop}%
\bibitem [{\citenamefont {Bravyi}\ and\ \citenamefont
  {Kitaev}(2002)}]{BRAVYI2002210}%
  \BibitemOpen
  \bibfield  {author} {\bibinfo {author} {\bibfnamefont {S.~B.}\ \bibnamefont
  {Bravyi}}\ and\ \bibinfo {author} {\bibfnamefont {A.~Y.}\ \bibnamefont
  {Kitaev}},\ }\href {\doibase https://doi.org/10.1006/aphy.2002.6254}
  {\bibfield  {journal} {\bibinfo  {journal} {Ann. Phys.}\ }\textbf {\bibinfo
  {volume} {298}},\ \bibinfo {pages} {210 } (\bibinfo {year}
  {2002})}\BibitemShut {NoStop}%
\bibitem [{\citenamefont {Seeley}\ \emph
  {et~al.}(2012{\natexlab{b}})\citenamefont {Seeley}, \citenamefont {Richard},\
  and\ \citenamefont {Love}}]{parity2012}%
  \BibitemOpen
  \bibfield  {author} {\bibinfo {author} {\bibfnamefont {J.~T.}\ \bibnamefont
  {Seeley}}, \bibinfo {author} {\bibfnamefont {M.~J.}\ \bibnamefont {Richard}},
  \ and\ \bibinfo {author} {\bibfnamefont {P.~J.}\ \bibnamefont {Love}},\
  }\href {\doibase 10.1063/1.4768229} {\bibfield  {journal} {\bibinfo
  {journal} {J. Chem. Phys.}\ }\textbf {\bibinfo {volume} {137}},\ \bibinfo
  {pages} {224109} (\bibinfo {year} {2012}{\natexlab{b}})}\BibitemShut
  {NoStop}%
\bibitem [{\citenamefont {Li}\ \emph {et~al.}(2019{\natexlab{b}})\citenamefont
  {Li}, \citenamefont {Hu}, \citenamefont {Zhang}, \citenamefont {Song},\ and\
  \citenamefont {Yung}}]{adts201800182}%
  \BibitemOpen
  \bibfield  {author} {\bibinfo {author} {\bibfnamefont {Y.}~\bibnamefont
  {Li}}, \bibinfo {author} {\bibfnamefont {J.}~\bibnamefont {Hu}}, \bibinfo
  {author} {\bibfnamefont {X.-M.}\ \bibnamefont {Zhang}}, \bibinfo {author}
  {\bibfnamefont {Z.}~\bibnamefont {Song}}, \ and\ \bibinfo {author}
  {\bibfnamefont {M.-H.}\ \bibnamefont {Yung}},\ }\href@noop {} {\bibfield
  {journal} {\bibinfo  {journal} {Adv. Theory and Simul.}\ }\textbf {\bibinfo
  {volume} {2}},\ \bibinfo {pages} {1800182} (\bibinfo {year}
  {2019}{\natexlab{b}})}\BibitemShut {NoStop}%
\bibitem [{\citenamefont {Dallaire-Demers}\ \emph {et~al.}(2018)\citenamefont
  {Dallaire-Demers}, \citenamefont {Romero}, \citenamefont {Veis},
  \citenamefont {Sim},\ and\ \citenamefont {Aspuru-Guzik}}]{Alan201801arxiv}%
  \BibitemOpen
  \bibfield  {author} {\bibinfo {author} {\bibfnamefont {P.-L.}\ \bibnamefont
  {Dallaire-Demers}}, \bibinfo {author} {\bibfnamefont {J.}~\bibnamefont
  {Romero}}, \bibinfo {author} {\bibfnamefont {L.}~\bibnamefont {Veis}},
  \bibinfo {author} {\bibfnamefont {S.}~\bibnamefont {Sim}}, \ and\ \bibinfo
  {author} {\bibfnamefont {A.}~\bibnamefont {Aspuru-Guzik}},\ }\href
  {https://arxiv.org/abs/1801.01053} {\bibfield  {journal} {\bibinfo  {journal}
  {arXiv:1801.01053}\ } (\bibinfo {year} {2018})}\BibitemShut {NoStop}%
\bibitem [{\citenamefont {Bonet-Monroig}\ \emph {et~al.}(2019)\citenamefont
  {Bonet-Monroig}, \citenamefont {Babbush},\ and\ \citenamefont
  {O'Brien}}]{RN818}%
  \BibitemOpen
  \bibfield  {author} {\bibinfo {author} {\bibfnamefont {X.}~\bibnamefont
  {Bonet-Monroig}}, \bibinfo {author} {\bibfnamefont {R.}~\bibnamefont
  {Babbush}}, \ and\ \bibinfo {author} {\bibfnamefont {T.~E.}\ \bibnamefont
  {O'Brien}},\ }\href@noop {} {\bibfield  {journal} {\bibinfo  {journal}
  {arXiv:1908.05628}\ } (\bibinfo {year} {2019})}\BibitemShut {NoStop}%
\bibitem [{\citenamefont {Yen}\ and\ \citenamefont {Izmaylov}(2020)}]{RN826}%
  \BibitemOpen
  \bibfield  {author} {\bibinfo {author} {\bibfnamefont {T.-C.}\ \bibnamefont
  {Yen}}\ and\ \bibinfo {author} {\bibfnamefont {A.~F.}\ \bibnamefont
  {Izmaylov}},\ }\href@noop {} {\bibfield  {journal} {\bibinfo  {journal}
  {arXiv:2007.01234}\ } (\bibinfo {year} {2020})}\BibitemShut {NoStop}%
\bibitem [{\citenamefont {Huang}\ and\ \citenamefont {Kueng}(2019)}]{RN824}%
  \BibitemOpen
  \bibfield  {author} {\bibinfo {author} {\bibfnamefont {R.-Y.}\ \bibnamefont
  {Huang}}\ and\ \bibinfo {author} {\bibfnamefont {i.}~\bibnamefont {Kueng}},\
  }\href@noop {} {\bibfield  {journal} {\bibinfo  {journal} {arXiv:1908.08909}\
  } (\bibinfo {year} {2019})}\BibitemShut {NoStop}%
\bibitem [{\citenamefont {Zhao}\ \emph {et~al.}(2020)\citenamefont {Zhao},
  \citenamefont {Tranter}, \citenamefont {Kirby}, \citenamefont {Ung},
  \citenamefont {Miyake},\ and\ \citenamefont {Love}}]{RN817}%
  \BibitemOpen
  \bibfield  {author} {\bibinfo {author} {\bibfnamefont {A.}~\bibnamefont
  {Zhao}}, \bibinfo {author} {\bibfnamefont {A.}~\bibnamefont {Tranter}},
  \bibinfo {author} {\bibfnamefont {W.~M.}\ \bibnamefont {Kirby}}, \bibinfo
  {author} {\bibfnamefont {S.~F.}\ \bibnamefont {Ung}}, \bibinfo {author}
  {\bibfnamefont {A.}~\bibnamefont {Miyake}}, \ and\ \bibinfo {author}
  {\bibfnamefont {P.~J.}\ \bibnamefont {Love}},\ }\href {\doibase
  10.1103/PhysRevA.101.062322} {\bibfield  {journal} {\bibinfo  {journal}
  {Phys. Rev. A}\ }\textbf {\bibinfo {volume} {101}},\ \bibinfo {pages}
  {062322} (\bibinfo {year} {2020})}\BibitemShut {NoStop}%
\bibitem [{\citenamefont {Huang}\ \emph {et~al.}(2020)\citenamefont {Huang},
  \citenamefont {Kueng},\ and\ \citenamefont {Preskill}}]{RN757}%
  \BibitemOpen
  \bibfield  {author} {\bibinfo {author} {\bibfnamefont {H.-Y.}\ \bibnamefont
  {Huang}}, \bibinfo {author} {\bibfnamefont {R.}~\bibnamefont {Kueng}}, \ and\
  \bibinfo {author} {\bibfnamefont {J.}~\bibnamefont {Preskill}},\ }\href@noop
  {} {\bibfield  {journal} {\bibinfo  {journal} {arXiv:2002.08953}\ } (\bibinfo
  {year} {2020})}\BibitemShut {NoStop}%
\bibitem [{\citenamefont {Yen}\ \emph {et~al.}(2020)\citenamefont {Yen},
  \citenamefont {Verteletskyi},\ and\ \citenamefont {Izmaylov}}]{RN784}%
  \BibitemOpen
  \bibfield  {author} {\bibinfo {author} {\bibfnamefont {T.-C.}\ \bibnamefont
  {Yen}}, \bibinfo {author} {\bibfnamefont {V.}~\bibnamefont {Verteletskyi}}, \
  and\ \bibinfo {author} {\bibfnamefont {A.~F.}\ \bibnamefont {Izmaylov}},\
  }\href {\doibase 10.1021/acs.jctc.0c00008} {\bibfield  {journal} {\bibinfo
  {journal} {J. Chem. Theory Comput.}\ }\textbf {\bibinfo {volume} {16}},\
  \bibinfo {pages} {2400} (\bibinfo {year} {2020})}\BibitemShut {NoStop}%
\bibitem [{\citenamefont {Izmaylov}\ \emph {et~al.}(2019)\citenamefont
  {Izmaylov}, \citenamefont {Yen}, \citenamefont {Lang},\ and\ \citenamefont
  {Verteletskyi}}]{RN626}%
  \BibitemOpen
  \bibfield  {author} {\bibinfo {author} {\bibfnamefont {A.~F.}\ \bibnamefont
  {Izmaylov}}, \bibinfo {author} {\bibfnamefont {T.-C.}\ \bibnamefont {Yen}},
  \bibinfo {author} {\bibfnamefont {R.~A.}\ \bibnamefont {Lang}}, \ and\
  \bibinfo {author} {\bibfnamefont {V.}~\bibnamefont {Verteletskyi}},\ }\href
  {https://doi.org/10.1021/acs.jctc.9b00791} {\bibfield  {journal} {\bibinfo
  {journal} {J. Chem. Theory Comput.}\ } (\bibinfo {year} {2019})}\BibitemShut
  {NoStop}%
\bibitem [{\citenamefont {Huggins}\ \emph {et~al.}(2019)\citenamefont
  {Huggins}, \citenamefont {McClean}, \citenamefont {Rubin}, \citenamefont
  {Jiang}, \citenamefont {Wiebe}, \citenamefont {Whaley},\ and\ \citenamefont
  {Babbush}}]{William2019arxiv}%
  \BibitemOpen
  \bibfield  {author} {\bibinfo {author} {\bibfnamefont {W.~J.}\ \bibnamefont
  {Huggins}}, \bibinfo {author} {\bibfnamefont {J.}~\bibnamefont {McClean}},
  \bibinfo {author} {\bibfnamefont {N.}~\bibnamefont {Rubin}}, \bibinfo
  {author} {\bibfnamefont {Z.}~\bibnamefont {Jiang}}, \bibinfo {author}
  {\bibfnamefont {N.}~\bibnamefont {Wiebe}}, \bibinfo {author} {\bibfnamefont
  {K.~B.}\ \bibnamefont {Whaley}}, \ and\ \bibinfo {author} {\bibfnamefont
  {R.}~\bibnamefont {Babbush}},\ }\href@noop {} {\bibfield  {journal} {\bibinfo
   {journal} {arXiv:1907.13117}\ } (\bibinfo {year} {2019})}\BibitemShut
  {NoStop}%
\bibitem [{\citenamefont {Abraham}\ and\ \citenamefont {et~al}(2019)}]{Qiskit}%
  \BibitemOpen
  \bibfield  {author} {\bibinfo {author} {\bibfnamefont {H.}~\bibnamefont
  {Abraham}}\ and\ \bibinfo {author} {\bibnamefont {et~al}},\ }\href {\doibase
  10.5281/zenodo.2562110} {\enquote {\bibinfo {title} {Qiskit: An open-source
  framework for quantum computing},}\ } (\bibinfo {year} {2019})\BibitemShut
  {NoStop}%
\bibitem [{\citenamefont {Morales}\ and\ \citenamefont {Nocedal}(2011)}]{bfgs}%
  \BibitemOpen
  \bibfield  {author} {\bibinfo {author} {\bibfnamefont {J.}~\bibnamefont
  {Morales}}\ and\ \bibinfo {author} {\bibfnamefont {J.}~\bibnamefont
  {Nocedal}},\ }\href@noop {} {\bibfield  {journal} {\bibinfo  {journal} {ACM
  Trans. Math. Softw.}\ }\textbf {\bibinfo {volume} {38}} (\bibinfo {year}
  {2011})}\BibitemShut {NoStop}%
\bibitem [{IBM()}]{IBM2020}%
  \BibitemOpen
  \href {https://quantum-computing.ibm.com} {\bibinfo  {journal}
  {https://quantum-computing.ibm.com}\ }\BibitemShut {NoStop}%
\bibitem [{\citenamefont {Stein}\ and\ \citenamefont
  {Reiher}(2016)}]{dmrg_active}%
  \BibitemOpen
\bibfield  {journal} {  }\bibfield  {author} {\bibinfo {author} {\bibfnamefont
  {C.~J.}\ \bibnamefont {Stein}}\ and\ \bibinfo {author} {\bibfnamefont
  {M.}~\bibnamefont {Reiher}},\ }\href {\doibase 10.1021/acs.jctc.6b00156}
  {\bibfield  {journal} {\bibinfo  {journal} {J. Chem. Theory Comput.}\
  }\textbf {\bibinfo {volume} {12}},\ \bibinfo {pages} {1760} (\bibinfo {year}
  {2016})}\BibitemShut {NoStop}%
\bibitem [{\citenamefont {Cincio}\ \emph
  {et~al.}(2021{\natexlab{a}})\citenamefont {Cincio}, \citenamefont {Rudinger},
  \citenamefont {Sarovar},\ and\ \citenamefont {Coles}}]{LZ_error2021}%
  \BibitemOpen
  \bibfield  {author} {\bibinfo {author} {\bibfnamefont {L.}~\bibnamefont
  {Cincio}}, \bibinfo {author} {\bibfnamefont {K.}~\bibnamefont {Rudinger}},
  \bibinfo {author} {\bibfnamefont {M.}~\bibnamefont {Sarovar}}, \ and\
  \bibinfo {author} {\bibfnamefont {P.~J.}\ \bibnamefont {Coles}},\ }\href
  {\doibase 10.1103/PRXQuantum.2.010324} {\bibfield  {journal} {\bibinfo
  {journal} {PRX Quantum}\ }\textbf {\bibinfo {volume} {2}},\ \bibinfo {pages}
  {010324} (\bibinfo {year} {2021}{\natexlab{a}})}\BibitemShut {NoStop}%
\bibitem [{\citenamefont {Czarnik}\ \emph {et~al.}(2020)\citenamefont
  {Czarnik}, \citenamefont {Arrasmith}, \citenamefont {Coles},\ and\
  \citenamefont {Cincio}}]{czarnik2020error}%
  \BibitemOpen
  \bibfield  {author} {\bibinfo {author} {\bibfnamefont {P.}~\bibnamefont
  {Czarnik}}, \bibinfo {author} {\bibfnamefont {A.}~\bibnamefont {Arrasmith}},
  \bibinfo {author} {\bibfnamefont {P.~J.}\ \bibnamefont {Coles}}, \ and\
  \bibinfo {author} {\bibfnamefont {L.}~\bibnamefont {Cincio}},\ }\href
  {https://arxiv.org/abs/2005.10189} {\bibfield  {journal} {\bibinfo  {journal}
  {arXiv:2005.10189}\ } (\bibinfo {year} {2020})}\BibitemShut {NoStop}%
\bibitem [{Note1()}]{Note1}%
  \BibitemOpen
  \bibinfo {note} {CDR training set is constructed using 4 near-Clifford
  circuits with 1 non-Clifford gate in each of 2 clusters. The training
  circuits are chosen by replacing non-Clifford gates with the closest Clifford
  gates.}\BibitemShut {Stop}%
\bibitem [{\citenamefont {Cincio}\ \emph
  {et~al.}(2021{\natexlab{b}})\citenamefont {Cincio}, \citenamefont {Rudinger},
  \citenamefont {Sarovar},\ and\ \citenamefont {Coles}}]{cincio2021machine}%
  \BibitemOpen
  \bibfield  {author} {\bibinfo {author} {\bibfnamefont {L.}~\bibnamefont
  {Cincio}}, \bibinfo {author} {\bibfnamefont {K.}~\bibnamefont {Rudinger}},
  \bibinfo {author} {\bibfnamefont {M.}~\bibnamefont {Sarovar}}, \ and\
  \bibinfo {author} {\bibfnamefont {P.~J.}\ \bibnamefont {Coles}},\ }\href
  {\doibase 10.1103/PRXQuantum.2.010324} {\bibfield  {journal} {\bibinfo
  {journal} {PRX Quantum}\ }\textbf {\bibinfo {volume} {2}},\ \bibinfo {pages}
  {010324} (\bibinfo {year} {2021}{\natexlab{b}})}\BibitemShut {NoStop}%
\bibitem [{\citenamefont {Ilya G.~Ryabinkin}()}]{iqcc2020}%
  \BibitemOpen
  \bibfield  {author} {\bibinfo {author} {\bibfnamefont {S.~N.~G.}\
  \bibnamefont {Ilya G.~Ryabinkin}, \bibfnamefont {Artur F.~Izmaylov}},\
  }\href@noop {} {\ }\BibitemShut {NoStop}%
\bibitem [{\citenamefont {Lange}\ and\ \citenamefont
  {Grubm{\"u}ller}(2008)}]{Lange2008-aq}%
  \BibitemOpen
  \bibfield  {author} {\bibinfo {author} {\bibfnamefont {O.~F.}\ \bibnamefont
  {Lange}}\ and\ \bibinfo {author} {\bibfnamefont {H.}~\bibnamefont
  {Grubm{\"u}ller}},\ }\href@noop {} {\bibfield  {journal} {\bibinfo  {journal}
  {Proteins}\ }\textbf {\bibinfo {volume} {70}},\ \bibinfo {pages} {1294}
  (\bibinfo {year} {2008})}\BibitemShut {NoStop}%
\bibitem [{\citenamefont {Lange}\ and\ \citenamefont
  {Grubm{\"u}ller}(2005)}]{Lange2005-if}%
  \BibitemOpen
  \bibfield  {author} {\bibinfo {author} {\bibfnamefont {O.~F.}\ \bibnamefont
  {Lange}}\ and\ \bibinfo {author} {\bibfnamefont {H.}~\bibnamefont
  {Grubm{\"u}ller}},\ }\href@noop {} {\enquote {\bibinfo {title} {Generalized
  correlation for biomolecular dynamics},}\ } (\bibinfo {year}
  {2005})\BibitemShut {NoStop}%
\bibitem [{\citenamefont {Rivalta}\ \emph {et~al.}(2012)\citenamefont
  {Rivalta}, \citenamefont {Sultan}, \citenamefont {Lee}, \citenamefont
  {Manley}, \citenamefont {Loria},\ and\ \citenamefont
  {Batista}}]{Rivalta2012-en}%
  \BibitemOpen
  \bibfield  {author} {\bibinfo {author} {\bibfnamefont {I.}~\bibnamefont
  {Rivalta}}, \bibinfo {author} {\bibfnamefont {M.~M.}\ \bibnamefont {Sultan}},
  \bibinfo {author} {\bibfnamefont {N.-S.}\ \bibnamefont {Lee}}, \bibinfo
  {author} {\bibfnamefont {G.~A.}\ \bibnamefont {Manley}}, \bibinfo {author}
  {\bibfnamefont {J.~P.}\ \bibnamefont {Loria}}, \ and\ \bibinfo {author}
  {\bibfnamefont {V.~S.}\ \bibnamefont {Batista}},\ }\href@noop {} {\bibfield
  {journal} {\bibinfo  {journal} {PNAS}\ }\textbf {\bibinfo {volume} {109}},\
  \bibinfo {pages} {E1428} (\bibinfo {year} {2012})}\BibitemShut {NoStop}%
\bibitem [{\citenamefont {Rogers}\ and\ \citenamefont
  {Singleton}(2020)}]{Rogers2020-fc}%
  \BibitemOpen
  \bibfield  {author} {\bibinfo {author} {\bibfnamefont {M.~L.}\ \bibnamefont
  {Rogers}}\ and\ \bibinfo {author} {\bibfnamefont {R.~L.}\ \bibnamefont
  {Singleton}},\ }\href@noop {} {\enquote {\bibinfo {title} {{Floating-Point}
  calculations on a quantum annealer: Division and matrix inversion},}\ }
  (\bibinfo {year} {2020})\BibitemShut {NoStop}%
\bibitem [{\citenamefont {Newman}(2006)}]{Newman2006}%
  \BibitemOpen
  \bibfield  {author} {\bibinfo {author} {\bibfnamefont {M.~E.~J.}\
  \bibnamefont {Newman}},\ }\href@noop {} {\bibfield  {journal} {\bibinfo
  {journal} {PNAS}\ }\textbf {\bibinfo {volume} {103}},\ \bibinfo {pages}
  {8577} (\bibinfo {year} {2006})}\BibitemShut {NoStop}%
\bibitem [{\citenamefont {Ushijima-Mwesigwa}\ \emph {et~al.}(2017)\citenamefont
  {Ushijima-Mwesigwa}, \citenamefont {Negre},\ and\ \citenamefont
  {Mniszewski}}]{Ushijima-Mwesigwa2017}%
  \BibitemOpen
  \bibfield  {author} {\bibinfo {author} {\bibfnamefont {H.}~\bibnamefont
  {Ushijima-Mwesigwa}}, \bibinfo {author} {\bibfnamefont {C.~F.~A.}\
  \bibnamefont {Negre}}, \ and\ \bibinfo {author} {\bibfnamefont {S.~M.}\
  \bibnamefont {Mniszewski}},\ }in\ \href@noop {} {\emph {\bibinfo {booktitle}
  {Proceedings of the Second International Workshop on Post Moores Era
  Supercomputing (PMES'17)}}}\ (\bibinfo  {publisher} {ACM New York, NY, USA},\
  \bibinfo {year} {2017})\ pp.\ \bibinfo {pages} {22--29}\BibitemShut {NoStop}%
\bibitem [{\citenamefont {Negre}\ \emph {et~al.}(2020)\citenamefont {Negre},
  \citenamefont {Ushijima-Mwesigwa},\ and\ \citenamefont
  {Mniszewski}}]{Negre2020}%
  \BibitemOpen
  \bibfield  {author} {\bibinfo {author} {\bibfnamefont {C.~F.~A.}\
  \bibnamefont {Negre}}, \bibinfo {author} {\bibfnamefont {H.}~\bibnamefont
  {Ushijima-Mwesigwa}}, \ and\ \bibinfo {author} {\bibfnamefont {S.~M.}\
  \bibnamefont {Mniszewski}},\ }\href@noop {} {\bibfield  {journal} {\bibinfo
  {journal} {PLOS ONE}\ }\textbf {\bibinfo {volume} {15}},\ \bibinfo {pages}
  {e0227538} (\bibinfo {year} {2020})}\BibitemShut {NoStop}%
\bibitem [{\citenamefont {Suzuki}(1990)}]{SUZUKI1990319}%
  \BibitemOpen
  \bibfield  {author} {\bibinfo {author} {\bibfnamefont {M.}~\bibnamefont
  {Suzuki}},\ }\href {\doibase https://doi.org/10.1016/0375-9601(90)90962-N}
  {\bibfield  {journal} {\bibinfo  {journal} {Phys. Lett. A}\ }\textbf
  {\bibinfo {volume} {146}},\ \bibinfo {pages} {319 } (\bibinfo {year}
  {1990})}\BibitemShut {NoStop}%
\bibitem [{\citenamefont {O'Malley}\ \emph {et~al.}(2016)\citenamefont
  {O'Malley}, \citenamefont {Babbush}, \citenamefont {Kivlichan}, \citenamefont
  {Romero}, \citenamefont {McClean}, \citenamefont {Barends}, \citenamefont
  {Kelly}, \citenamefont {Roushan}, \citenamefont {Tranter}, \citenamefont
  {Ding}, \citenamefont {Campbell}, \citenamefont {Chen}, \citenamefont {Chen},
  \citenamefont {Chiaro}, \citenamefont {Dunsworth}, \citenamefont {Fowler},
  \citenamefont {Jeffrey}, \citenamefont {Lucero}, \citenamefont {Megrant},
  \citenamefont {Mutus}, \citenamefont {Neeley}, \citenamefont {Neill},
  \citenamefont {Quintana}, \citenamefont {Sank}, \citenamefont {Vainsencher},
  \citenamefont {Wenner}, \citenamefont {White}, \citenamefont {Coveney},
  \citenamefont {Love}, \citenamefont {Neven}, \citenamefont {Aspuru-Guzik},\
  and\ \citenamefont {Martinis}}]{PhysRevX.6.031007}%
  \BibitemOpen
  \bibfield  {author} {\bibinfo {author} {\bibfnamefont {P.~J.~J.}\
  \bibnamefont {O'Malley}}, \bibinfo {author} {\bibfnamefont {R.}~\bibnamefont
  {Babbush}}, \bibinfo {author} {\bibfnamefont {I.~D.}\ \bibnamefont
  {Kivlichan}}, \bibinfo {author} {\bibfnamefont {J.}~\bibnamefont {Romero}},
  \bibinfo {author} {\bibfnamefont {J.~R.}\ \bibnamefont {McClean}}, \bibinfo
  {author} {\bibfnamefont {R.}~\bibnamefont {Barends}}, \bibinfo {author}
  {\bibfnamefont {J.}~\bibnamefont {Kelly}}, \bibinfo {author} {\bibfnamefont
  {P.}~\bibnamefont {Roushan}}, \bibinfo {author} {\bibfnamefont
  {A.}~\bibnamefont {Tranter}}, \bibinfo {author} {\bibfnamefont
  {N.}~\bibnamefont {Ding}}, \bibinfo {author} {\bibfnamefont {B.}~\bibnamefont
  {Campbell}}, \bibinfo {author} {\bibfnamefont {Y.}~\bibnamefont {Chen}},
  \bibinfo {author} {\bibfnamefont {Z.}~\bibnamefont {Chen}}, \bibinfo {author}
  {\bibfnamefont {B.}~\bibnamefont {Chiaro}}, \bibinfo {author} {\bibfnamefont
  {A.}~\bibnamefont {Dunsworth}}, \bibinfo {author} {\bibfnamefont {A.~G.}\
  \bibnamefont {Fowler}}, \bibinfo {author} {\bibfnamefont {E.}~\bibnamefont
  {Jeffrey}}, \bibinfo {author} {\bibfnamefont {E.}~\bibnamefont {Lucero}},
  \bibinfo {author} {\bibfnamefont {A.}~\bibnamefont {Megrant}}, \bibinfo
  {author} {\bibfnamefont {J.~Y.}\ \bibnamefont {Mutus}}, \bibinfo {author}
  {\bibfnamefont {M.}~\bibnamefont {Neeley}}, \bibinfo {author} {\bibfnamefont
  {C.}~\bibnamefont {Neill}}, \bibinfo {author} {\bibfnamefont
  {C.}~\bibnamefont {Quintana}}, \bibinfo {author} {\bibfnamefont
  {D.}~\bibnamefont {Sank}}, \bibinfo {author} {\bibfnamefont {A.}~\bibnamefont
  {Vainsencher}}, \bibinfo {author} {\bibfnamefont {J.}~\bibnamefont {Wenner}},
  \bibinfo {author} {\bibfnamefont {T.~C.}\ \bibnamefont {White}}, \bibinfo
  {author} {\bibfnamefont {P.~V.}\ \bibnamefont {Coveney}}, \bibinfo {author}
  {\bibfnamefont {P.~J.}\ \bibnamefont {Love}}, \bibinfo {author}
  {\bibfnamefont {H.}~\bibnamefont {Neven}}, \bibinfo {author} {\bibfnamefont
  {A.}~\bibnamefont {Aspuru-Guzik}}, \ and\ \bibinfo {author} {\bibfnamefont
  {J.~M.}\ \bibnamefont {Martinis}},\ }\href {\doibase
  10.1103/PhysRevX.6.031007} {\bibfield  {journal} {\bibinfo  {journal} {Phys.
  Rev. X}\ }\textbf {\bibinfo {volume} {6}},\ \bibinfo {pages} {031007}
  (\bibinfo {year} {2016})}\BibitemShut {NoStop}%
\end{thebibliography}%

\end{document}